\UseRawInputEncoding
\documentclass[aps,prb,reprint,amsmath, amssymb, aps,superscriptaddress]{revtex4-2}

\usepackage[dvips]{graphicx,color}
\usepackage{dcolumn}
\usepackage{bm}
\usepackage[]{lineno}

\def \k {{\bm k}}
\def \HH {{\textrm{HH}}}
\def \TT {{\textrm{TT}}}
\def \HT {{\textrm{HT}}}

\begin{document}
%

\title{Nodal-line semimetal HMTSF-TCNQ: Anomalous orbital diamagnetism and charge density wave}


\author{S. Ozaki}
\email[]{ozaki@hosi.phys.s.u-tokyo.ac.jp}
\affiliation{Department of Physics, University of Tokyo, Bunkyo, Tokyo 113-0033, Japan}
\author{I. Tateishi}
\affiliation{Department of Physics, University of Tokyo, Bunkyo, Tokyo 113-0033, Japan}
\affiliation{RIKEN Center for Emergent Matter Science, Wako, Saitama 351-0198, Japan}
\author{H. Matsuura}
\affiliation{Department of Physics, University of Tokyo, Bunkyo, Tokyo 113-0033, Japan}
\author{M. Ogata}
\affiliation{Department of Physics, University of Tokyo, Bunkyo, Tokyo 113-0033, Japan}
\affiliation{Trans-scale Quantum Science Institute, University of Tokyo, Bunkyo, Tokyo 113-0033, Japan}
\author{K. Hiraki}
\affiliation{Fukushima Medical University, Fukushima 960-1295, Japan}
\affiliation{IMR, Tohoku University, Miyagi 980-8577, Japan}


\date{\today}

\begin{abstract}
This study investigates the electronic states and physical quantities of an organic charge-transfer complex 
HMTSF-TCNQ, which undergoes 
a charge-density-wave (CDW) phase transition at temperature $T_c\simeq30$ K.
A first-principles calculation is utilized to determine that
the normal state is a topological semimetal with open nodal lines.
Besed on the first-principles calculation, we develop a tight-binding model
to investigate the electronic state in detail.
Below $T_c$, the CDW phase is examined in the tight-binding scheme 
using the mean-field approximation. It is shown 
that the open nodal lines are deformed into closed ones,
and their shapes are sensitive to the order parameter.
Using this tight-binding model, we theoretically evaluate the temperature dependencies of two physical quantities:
the spin-lattice relaxation time $T_1$ and the orbital magnetic susceptibility.
In particular, an anomalous plateau is obtained at low temperatures in the orbital diamagnetism.
We presume that this anomalous plateau originates owing to the conflict between 
the interband diamagnetism, impurity scattering, and the nodal line deformation.
We also conduct an experiment to investigate the orbital magnetism,
and the results are in excellent quantitative agreement with the theory.
\end{abstract}


\maketitle


\section{introduction\label{sec:intro}}
A charge density wave (CDW) in topological bands induces fascinating phenomena,
such as the three-dimensional Hall effect \cite{3dhall} and axionic CDW phase \cite{axionic},
which have recently generated considerable research interest.
A CDW is a quantum phase that is typical of quasi-one-dimensional organic conductors,
and the topological properties of organic materials have become more extensively researched  
in the last two decades \cite{kobayashi07,kobayashi08,wang-prl,zhang-nanoletter,liu-prb,kato-jacs,
kato-jpsj, kawamura-jpsj,kato2020}.
Organic materials that possess topological properties 
have the potential for novel phenomena that originate from the interplay between 
their topological bands and the CDW.
Such materials require both one-dimensionality for the CDW and 
two- or three-dimensionality for the topological nature.
One example of these materials is HMTSF-TCNQ (hexamethylene-tetraselena-fulvalene-tetracyanoquindimethane)
\cite{bloch75,bechgaard,soda,weger}.

HMTSF-TCNQ is a classical quasi-one-dimensional charge-transfer complex 
that was discovered in the 1970s,
and it has attracted considerable attention in contemporary research,
owing to the possibility of field-induced CDW 
\cite{murata-lowtemp,murata-physica,murata-jpsj}.
HMTSF-TCNQ has components that are similar to those of the well-known organic complexes TTF-TCNQ and
TMTSF$_2$PF$_6$, which are known as a typical organic conductor that has a CDW and 
the first organic superconductor identified, respectively \cite{jerome-review,jerome04}.
These two materials have been studied extensively; however,
HMTSF-TCNQ has not.

HMTSF-TCNQ undergoes a CDW transition at approximately 30 K under ambient pressure; however,
the temperature dependence of the resistivity is different from that of TTF-TCNQ 
in that a clear metal-insulator transition is not observed.
According to Refs.~\cite{jerome-review,weger}, the Fermi surface of HMTSF-TCNQ is two-dimensional,
owing to the relatively large interchain hoppings,
which causes incomplete CDW nesting.
In this material, unconventional temperature dependencies of physical quantities, such as 
the magnetic susceptibility \cite{soda},
Seebeck coefficient \cite{bloch75}, and the electric conductivity are known
and are presumed to be attributed to the incomplete CDW nesting.

Despite these intriguing properties, there are no reliable theoretical models
for the electronic states of the normal phase or the CDW phase.
The electronic state of the normal phase is typically discussed in terms of
Weger's model \cite{weger},
which is a four-band tight-binding model that consists of 
HOMOs (highest occupied molecular orbitals) and 
LUMOs (lowest unoccupied molecular orbitals) for 
each molecule on the basis of the crystal structure known at that time 
with the estimated hopping parameters. 
This model was able to explain some experimental results;
however, a subsequent experimental study suggested the crystal structure of a different space group
\cite{phillips}.
Furthermore, the first-principles calculation based on the new crystal structure,
which is presented in this study,
suggests Fermi surfaces of different shapes and locations
from those calculated by Weger.
Therefore, we need to modify the existing model on the basis of these results.

The physical quantities of HMTSF-TCNQ are of interest as well as its electronic states.
In particular, HMTSF-TCNQ shows a large diamagnetism at low temperatures, and 
it reaches an anomalous plateau below $T_c$ \cite{soda}.
Weger tried to explain this behavior via the model and concluded that 
the diamagnetism was attributed to Landau diamagnetism.
However, this claim is debatable because Landau diamagnetism is derived from the electrons on the Fermi surface.
In the CDW state, the density of states (DOS) at the Fermi level is expected to be small.
Therefore, Landau diamagnetism does not explain the large diamagnetism at low temperatures; thus,
the origin of the diamagnetism remains unknown.

To address this problem, the current study serves a twofold purpose.
One is to reveal the electronic states of the normal and CDW phases of HMTSF-TCNQ 
under ambient pressure.
First, we implement a first-principles calculation based on the correct crystal structure.
The obtained results, in particular the symmetry indicator, 
indicate that its normal phase is a topologically protected nodal line semimetal.
On the basis of these results, we construct a tight-binding model 
and discuss the nodal lines in detail.
These analyses clarify that 
the energy at the node points fluctuate along the nodal lines, which results in 
electron and hole pockets.
As a result, we expect that the nodal lines influence various physical quantities that are susceptible to 
the Fermi surface.
Furthermore, assuming a plausible CDW nesting, we discuss how the CDW affects
the nodal lines and the electron and hole pockets.
We theoretically evaluate the spin-lattice relaxation time $T_1$ 
for the experimental confirmation of the existence of the nodal lines.

The other objective is to clarify the derivation of the large diamagnetism and 
its temperature dependence.
Since the nodal lines are located near the Fermi level, 
a large diamagnetism derived from the interband effect is strongly expected
from an analogy to the case of the Dirac electrons 
\cite{fukuyama-kubo, fukuyama07,nakamura07,koshino-ando,raoux-piechon,fuseya,ozaki-ogata}. 
Note that the interband effect is insensitive to the DOS at the Fermi level.
Using the Fukuyama formula \cite{fukuyama71}, which properly describes the interband effect,
we evaluate the magnetic susceptibility.
An experiment is conducted to determine the magnetic susceptibility using 
a superconducting quantum interference device (SQUID) magnetometer. 
The proposed theory is in excellent agreement with this experimental result.

This paper is organized as follows. 
In Sec.~\ref{sec:structure},
the correct crystal structure of HMTSF-TCNQ is presented.
In Sec.~\ref{sec:firstprinciple}, we investigate the electronic state of HMTSF-TCNQ 
using the first-principles calculation, and we demonstrate the existence of nodal lines.
In Sec.~\ref{sec:tightbinding}, we construct a tight-binding model 
based on the results of the previous section and discuss the nodal lines in detail.
In Sec.~\ref{sec:cdw}, we introduce the CDW with the mean-field approximation and examine the 
CDW phase.
In Sec.~\ref{sec:physq}, we evaluate the spin-lattice relaxation time $T_1$ and 
the orbital magnetic susceptibility. 
A comparison between the theory and the magnetic susceptibility experimental results
is also conducted.
Finally, we present the conclusion to the study in Sec.~\ref{sec:summary}.

\section{Crystal structure of HMTSF-TCNQ
\label{sec:structure}}
\begin{figure*}[t]
	\begin{center}
	\includegraphics[width=17.8cm]{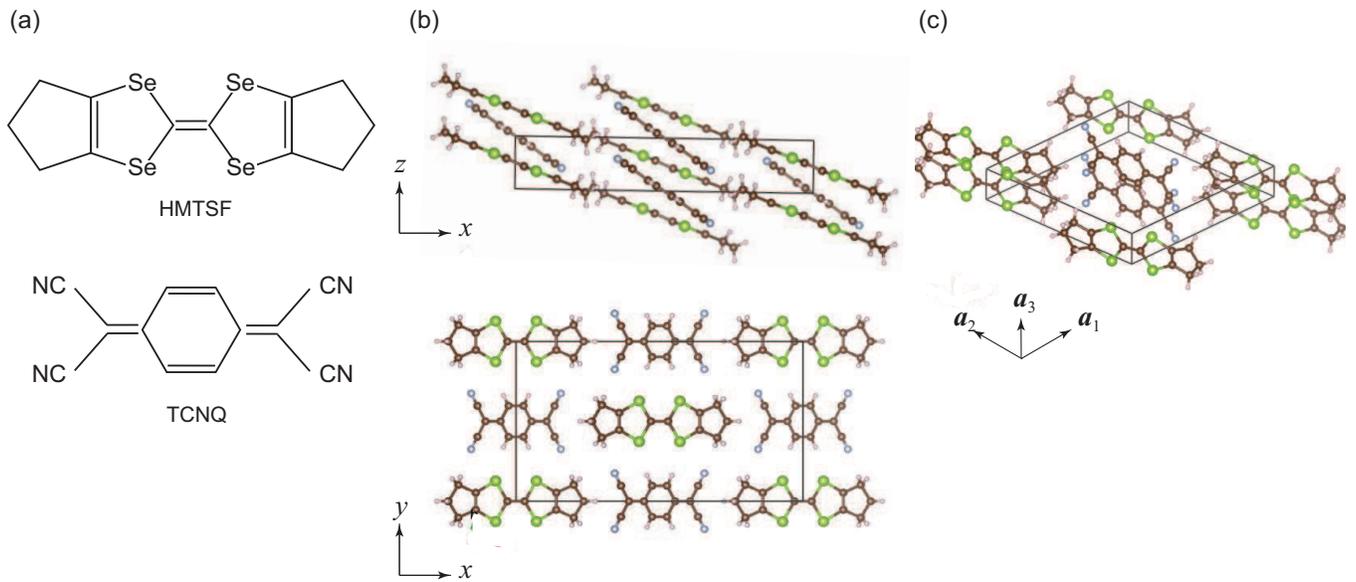}
	\caption{(a) Molecular structures of HMTSF and TCNQ.
	(b) Conventional and (c) primitive cell of HMTSF-TCNQ.
	Brown, green, blue, and white circles represent 
	C, Se, N, and H atoms, respectively.
	The black lines represent the boundary of each unit cell.
	The primitive cell is half the size of the conventional cell.
	\label{fig:molecules}}
	\end{center}
\end{figure*}
\begin{figure}[h]
\begin{center}
\includegraphics[width=\linewidth]{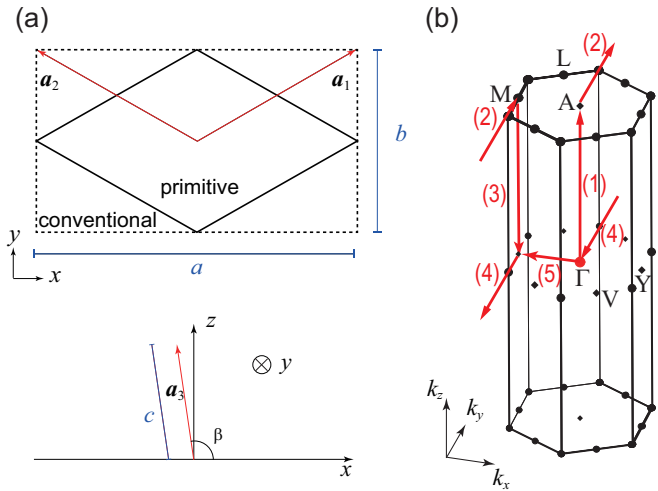}
\caption{(a) Basic lattice vectors for the primitive cell of HMTSF-TCNQ.
The rhombus and dashed rectangle represent the boundary of 
the primitive and conventional cell, respectively.
The lattice constants $a$, $b$, $c$, and $\beta$ are for the conventional cell.
(b) First Brillouin zone and TRIM. 
The TRIM are $\Gamma$, A, Y, M, two V, and two L points.
The path for the energy dispersion is also shown.}
\label{fig:primitivevector}
\end{center}
\end{figure}

HMTSF-TCNQ consists of two organic molecules, HMTSF and TCNQ, 
the structure of which are shown in Fig.~\ref{fig:molecules}(a).
This material has a base-centered monoclinic lattice,
and for convenience, we introduce two different notations of the unit cell:
the conventional cell and the primitive cell,
which are shown in Figs.~\ref{fig:molecules}(b) and \ref{fig:molecules}(c), respectively.
The crystal structures and wave functions in this study are drawn using VESTA (JP minerals, Japan) \cite{vesta}.
Each molecule forms a one-dimensional chain, and the HMTSF and TCNQ chains are arranged
in a checkerboard configuration, as can be observed in Fig.~\ref{fig:molecules}(b).
Throughout this study, we use the Cartesian coordinate system, as shown in 
Fig.~\ref{fig:primitivevector}(a)
(the unique axis ${\bm b}$ convention).
The space group is $C2/m$ (No.~12), the generators of which are lattice translations, $C_{2y}$ rotation, and inversion $I$
\cite{hahn}.
Recent measurements for high quality samples have determined the lattice constants 
and atomic coordinates at room temperature \cite{kato}, which are used in this study.
The experimental details will be presented elsewhere.
The newly obtained lattice constants are slightly different from those 
presented in a previous study \cite{phillips}.
These values are tabulated in Table~\ref{table:latticeconstants}.

Figure~\ref{fig:primitivevector}(a) shows the basic lattice vectors for the primitive cell,
${\bm a}_1=(a/2, b/2, 0)$, ${\bm a}_2=(-a/2, b/2, 0)$, 
and ${\bm a}_3=(c\cos \beta, 0, c\sin\beta)$.
The orientation of the chains are parallel to ${\bm a}_3$,
and the corresponding reciprocal lattice vectors are 
${\bm b_1}=2\pi(1/a, 1/b, -1/a\tan\beta)$, 
${\bm b_2}=2\pi(-1/a, 1/b, 1/a\tan\beta)$, and
${\bm b_3}=2\pi(0,0,1/c\sin\beta)$.
Note that ${\bm a_3}$ is almost parallel to the $z$ axis, but slightly inclined towards the negative $x$ direction
and that ${\bm b_3}$ is parallel to the $k_z$ axis.
Figure \ref{fig:primitivevector}(b) shows the first Brillouin zone,
which is a hexagonal column, and the time-reversal invariant momenta (TRIM) are
$\Gamma$, A, Y, M, two V, and two L points.

\begin{table}[h]
	\caption{Lattice constants for the conventional cell.}
   \label{table:latticeconstants}
	 \begin{ruledtabular}
		\begin{tabular}{ccccc}
  		 & $a$/\AA & $b$/\AA& $c$/\AA& 
  		$\beta/^\circ$  \\
			\colrule
			Ref.~\cite{phillips} & 21.999(14) & 12.573 & 3.890(1) & 90.29(4) \\
  		present \cite{kato} & 21.85 & 12.48 & 3.87 & 90.25 \\
  	\end{tabular}
	\end{ruledtabular}
\end{table}

\section{First-principles calculation of HMTSF-TCNQ \label{sec:firstprinciple}}
In this section, we present the electronic state of HMTSF-TCNQ, obtained by a first-principles calculation. 
This calculation is performed by QUANTUM ESPRESSO \cite{qe},
which uses the density functional theory \cite{hohenberg-kohn, kohn-sham}.
We neglect spin-orbit interaction because in organic materials,
it is typically negligible.
For the exchange-correlation term, the generalized gradient approximation 
with nonrelativistic Perdew-Burke-Ernzerhof parametrization \cite{pbe96} 
is used.
The Kohn-Sham orbitals are expanded with plane waves and the cutoff 
energies are 70 and 320 Ry for wave functions and charge density, respectively.
The $\k$-point grid on the Brillouin zone is taken as $8\times 8 \times 40$.

\begin{figure*}
\begin{center}
\includegraphics[width=\linewidth]{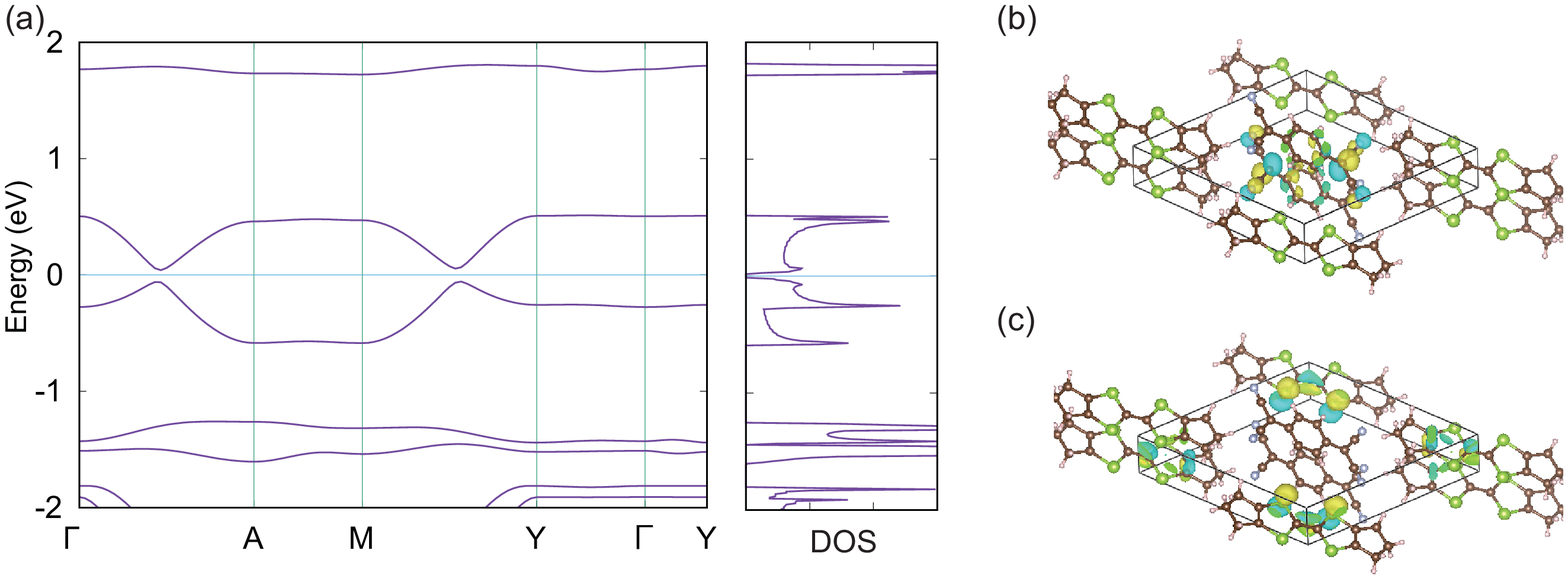}
\caption{(a) Energy dispersion and DOS of HMTSF-TCNQ.
Wave functions for (b) valence and (c) conduction bands at the $\Gamma$ point.
These wave functions consist of two Wannier functions,
which are the same for wave functions at other high-symmetry points,
shown in Appendix~\ref{sec:wavefunctions}.}
\label{fig:dft}
\end{center}
\end{figure*}
The obtained energy dispersion and DOS are shown in Fig.~\ref{fig:dft}(a),
and the correspoding path is shown in Fig.~\ref{fig:primitivevector}(b).
First, we find that the dispersion is almost flat in the $k_x$ and $k_y$ directions,
and the DOS has peaks at the band edges.
These properties are characteristic of quasi-one-dimensional materials.
Another important feature is that the two bands near the Fermi level
are very isolated from other bands.
This implies that the mixing of other orbitals is small and 
that it is sufficient to consider these two bands to discuss the 
low-energy properties of this material.
An energy gap is observed in the energy dispersion at the Fermi energy in Fig.~\ref{fig:dft}(a); however,
the DOS does not show a clear gap.
As shown below, this is because there are nodal lines away from the high-symmetry ${\bm k}$ points.
This Dirac-type dispersion may strongly affect physical quantities,
such as the orbital diamagnetism.

Next, we examine the spatial distribution of the wave functions.
Figures~\ref{fig:dft}(b) and \ref{fig:dft}(c) show the wave functions at the $\Gamma$ point
for the valence and conduction bands, respectively.
We observe that the wave functions at other high-symmetry points also
consist of these two Wannier functions.
(The wave functions at other high-symmetry points are shown in Appendix \ref{sec:wavefunctions}.)
Therefore, we can naturally assume that the two bands at each $\k$ point are well described 
by the linear combination of these two Wannier functions.

Thirdly, we discuss the topological nature of the material using the symmetry indicator.
We can observe that there is no band crossing on the high-symmetry lines or planes.
However, the space group No.~12 can have nodal lines at generic points 
\cite{fu-kane,pvw,song-natcom,youngkuk}.
The existence of nodal lines is diagnosed by the symmetry indicator $(z_{2,2}, z'_2)$,
the components of which are defined by
\begin{linenomath}
\begin{align}
	z_{2,2}&=\sum_{\k=\mathrm{V,Y,M,L}} n^-(\k) ~~(\mathrm{mod}~2), \\
	z'_2&=\frac{1}{2}\sum_{\k=8\mathrm{TRIM}} n^-(\k) ~~(\mathrm{mod}~2),
\end{align}
\end{linenomath}
where $n^-(\k)$ is the number of occupied bands with odd parity at $\k$.
This indicator is calculated by examining the parities of the wave functions, 
and we obtain a nontrivial indicator  $(z_{2,2},z'_2)=(1,1)$.
This value indicates the existence of open nodal lines approximately along the $k_x$ direction \cite{song-prx}.
(The parities at TRIM are shown in Appendix \ref{sec:wavefunctions}.)

\section{Tight-binding model \label{sec:tightbinding} and nodal lines}
\subsection{Tight-binding model}
In the previous section, 
the discussion based on the symmetry indicator suggests the existence of the nodal lines.
However, the symmetry indicator does not indicate their exact locations or detailed properties.
Therefore, we construct a tight-binding model and 
analyze it to clarify the detailed electronic properties.

\begin{figure}
	\begin{center}
	\includegraphics[width=\linewidth]{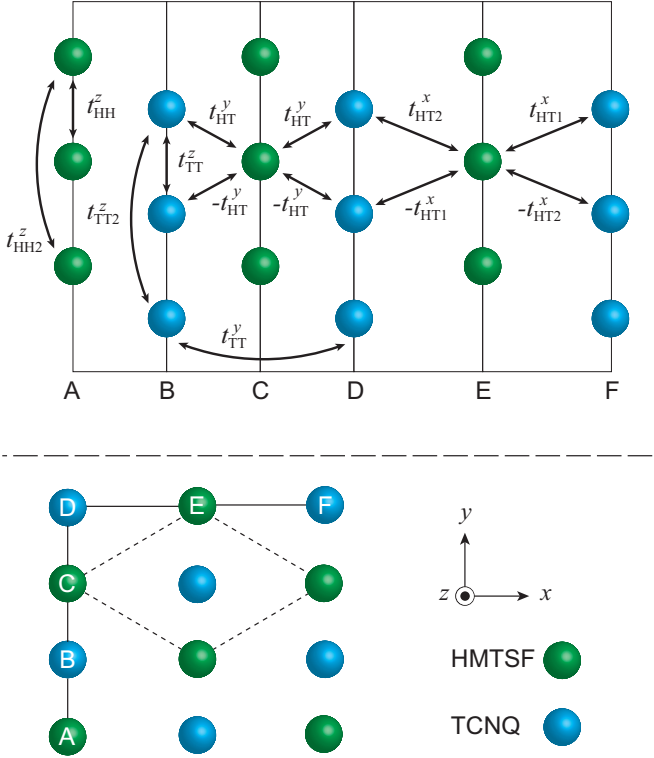}
	\caption{Eight largest hopping parameters used in the tight-binding model [Eq.~(\ref{tbham})].
	The subscript and superscript represent the related molecules and the direction of the hopping,
	respectively.
	The lower panel shows the configuration of the molecules viewed from the $z$ direction,
	and A--F indicate the positions of atoms in the upper panel.
	The dashed line represents the primitive unit cell.}
	\label{fig:hoppings}
	\end{center}
\end{figure}

As shown in the previous section, the wave functions near the Fermi level
consist of the two Wannier functions located in HMTSF and TCNQ molecules.
From these orbitals, we construct a two-band tight-binding model using Slater-Koster's method \cite{sk}.
We fit the hopping parameters to reproduce the band dispersion using WANNIER90 \cite{wannier90}. 
From this result, 
we consider the eight largest hopping parameters and neglect the others.
The chosen hoppings are shown in Fig.~\ref{fig:hoppings}, and 
the fitted parameters are summarized in Table~\ref{table:hoppings}.
We measure the energies from the Fermi level (i.e., $\mu=0$) and denote the one-body level for 
the HMTSF (TCNQ) orbital as $\varepsilon_{\rm H}$ ($\varepsilon_{\rm T}$).
Using the Fourier transform, we obtain the two-band tight-binding model in the ${\bm k}$ space as follows:
\begin{linenomath}
\begin{equation}
	H_0(\k)=
	\left(
	\begin{array}{cc}
		t_{\HH} ({\bm k}) & t_{\HT}^* ({\bm k}) \\
		t_{\HT}({\bm k}) & t_{\TT} ({\bm k}) \\
	\end{array}
	\right),
	\label{eq:tbmat}
	\end{equation}
\end{linenomath}
where
\begin{linenomath}
\begin{align}
	t_{\HH}(\k)=&\varepsilon_{\rm H}+ 2t_{\HH}^z \cos \tilde{k}_z c 
	+ 2 t_{\HH 2}^z \cos 2\tilde{k}_z c + 2t_{\HH}^y \cos k_y b, \\
	t_{\TT}(\k)=&\varepsilon_{\rm T}+ 2t_{\TT}^z \cos \tilde{k}_z c
	+ 2 t_{\TT 2}^z \cos 2\tilde{k}_z c + 2t_{\TT}^y \cos k_y b, \\
	t_{\HT}(\k)=&4i t_{\HT}^y \cos \frac{k_y b}{2} \sin \frac{\tilde{k}_z c}{2}
	+ 2it_{\HT 1}^x \sin \left(-\frac{k_x a}{2} +\frac{\tilde{k}_z c}{2}\right) \nonumber \\
	&+2it_{\HT 2}^x \sin \left(\frac{k_x a}{2} +\frac{\tilde{k}_z c}{2}\right),
	\label{tbham}
\end{align}
\end{linenomath}
with $\tilde{k}_z=k_x \cos\beta + k_z\sin\beta $.

\begin{table}[hb]
	\caption{Hopping parameteres between Wannier functions.}
	\label{table:hoppings}
	\begin{ruledtabular}
	\begin{tabular}{cd}
	\textrm{Hopping parameters} &
		\multicolumn{1}{c}{\textrm{Energy / eV}} \\ \colrule
		$t_{\HH}^z$ &  0.269 \\
		$t_{\TT}^z$ & -0.184 \\
		$t_{\HH 2}^z$ & 0.0106 \\
		$t_{\TT 2}^z$ & 0.0126 \\
		$t_{\TT}^y$ & -0.00358 \\
		$t_{\HT}^y$ & 0.0193 \\
		$t_{\HT 1}^x$ & -0.00689 \\ 
		$t_{\HT 2}^x$ & 0.00446 \\  
		\colrule
     $\varepsilon_{\rm H}$ & -0.0316 \\  
     $\varepsilon_{\rm T}$ &  0.0542 \\ 
	\end{tabular}
	\end{ruledtabular}
\end{table}

\begin{figure*}
\begin{center}
\includegraphics[width=\linewidth]{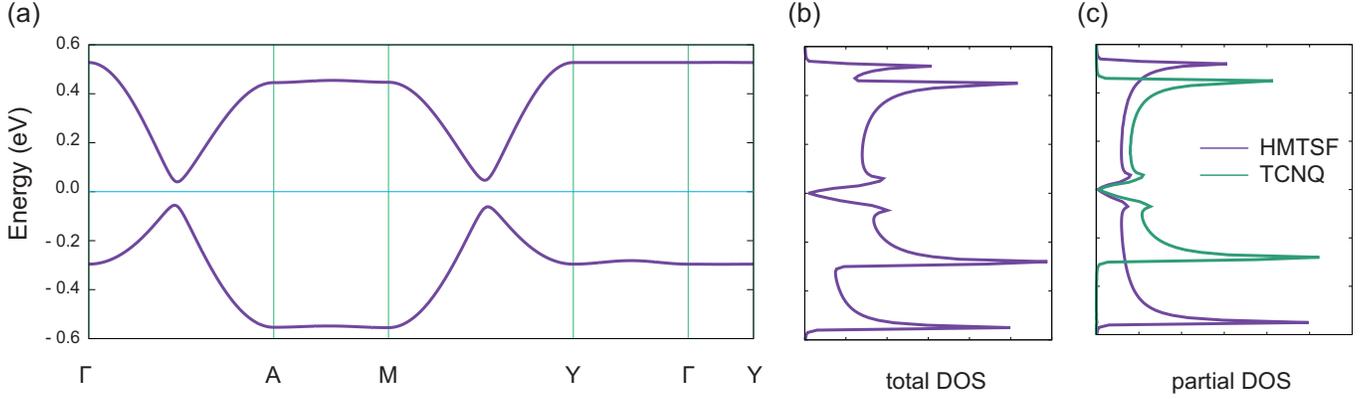}
\caption{(a) Energy dispersion and (b) the total DOS of the proposed tight-binding model. 
(c) Partial DOS for HMTSF and TCNQ.}
\label{fig:banddos-tb}
\end{center}
\end{figure*}

Diagonalizing the Hamiltonian in Eq.~(\ref{eq:tbmat}), we obtain the energy dispersion:
\begin{linenomath}
\begin{equation}
	E_\pm = \frac{t_{\HH}(\k)+t_{\TT}(\k)}{2}
	\pm \sqrt{|t_{\HT}(\k)|^2 + \frac{1}{4}(t_{\HH}(\k)-t_{\TT}(\k))^2},
	\label{eq:energydisp}
\end{equation} 
\end{linenomath}
which is shown in Fig.~\ref{fig:banddos-tb}(a).
The total DOS is shown in Fig.~\ref{fig:banddos-tb}(b).
These results accurately reproduce those 
by the first-principles calculation shown in Fig.~\ref{fig:dft}(a).
In the DOS, we observe a linear behavior near the Fermi level, which is typical for two-dimensional Dirac electrons. 
Figure~\ref{fig:banddos-tb}(c) shows the partial DOS for HMTSF and TCNQ molecules.
While these wave functions decouple at the band edge, 
they contribute almost equally to the DOS at the Fermi level.
This result supports the existence of the Dirac-type dispersion,
because it always consists of at least two orbitals.
The shape of the Dirac cone calculated from the proposed tight-binding model 
is shown in Appendix~\ref{sec:diraccone}.

\subsection{Nodal lines}
\begin{figure*}
\begin{center}
\includegraphics[width=\linewidth]{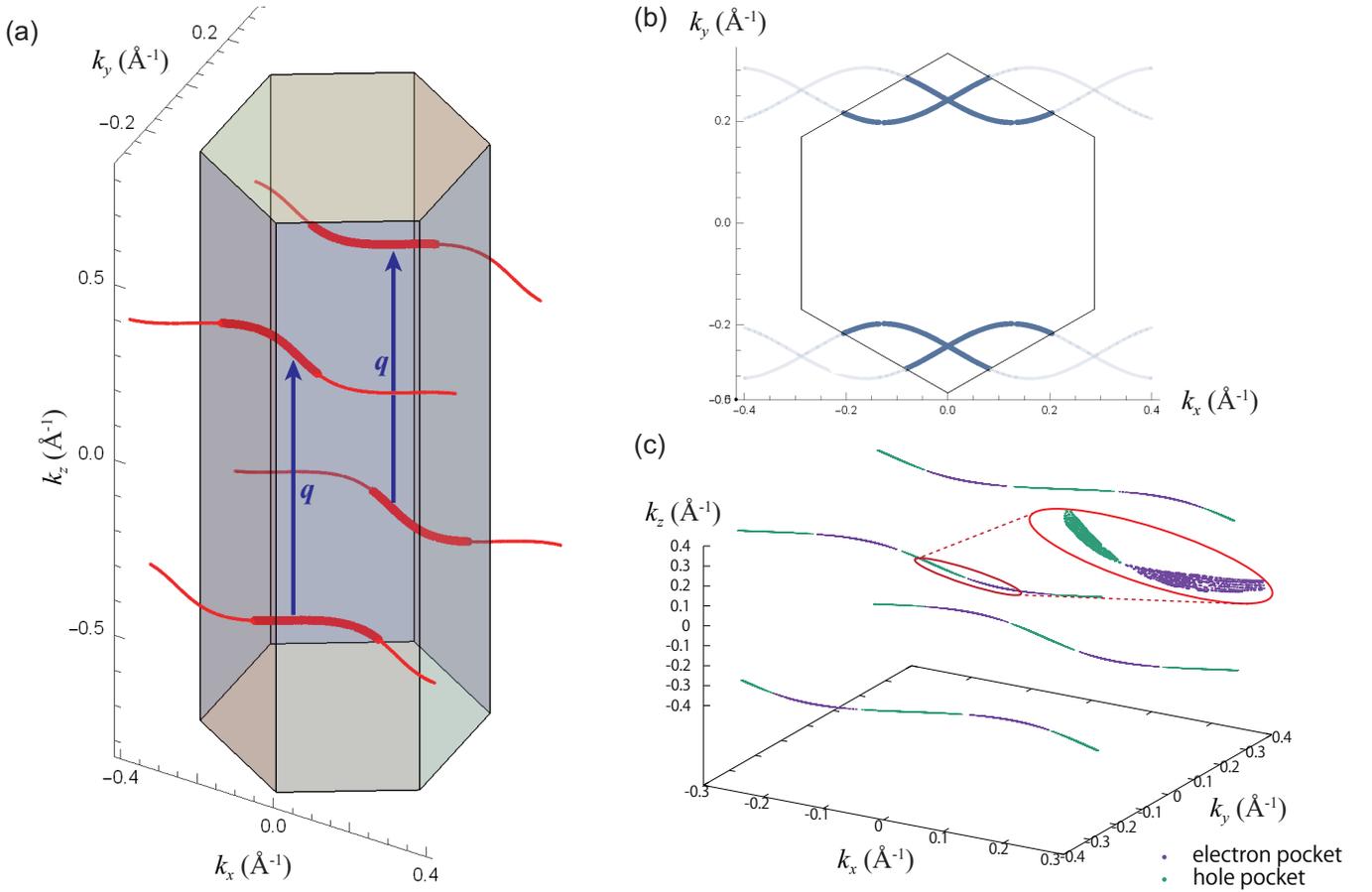}
\caption{(a) Configuration of the nodal lines in the ${\bm k}$ space. 
The nodal lines in (out of) the first Brillouin zone are indicated by thick (thin) lines.
The nodal lines have $C_{2y}$ rotation, inversion, and $\sigma_y$ mirror symmetry.
The vector ${\bm q}$ represents the CDW nesting vector to be considered in Sec.~\ref{sec:cdw},
which is on the $k_x=0$ plane.
(b) Nodal lines projected onto the $k_x$-$k_y$ plane.
They intersect on the $k_x=0$ line, owing to the $C_{2y}$ rotational symmetry.
(c) Fermi surface of HMTSF-TCNQ.
Purple and green surfaces represent the electron and hole pockets, respectively.
The Fermi surface consists of the alternating electron and hole pockets
along the nodal lines.
\label{fig:nodal}
}
\end{center}
\end{figure*}

This subsection discusses the locations of the nodal lines in the ${\bm k}$ space.
From the energy dispersion Eq.~(\ref{eq:energydisp}), 
we can identify the nodal lines by the following condition:
\begin{linenomath}
\begin{equation}
	t_{\HT}(\k)=0, \quad t_{\HH}(\k)=t_{\TT}(\k). \label{eq:nodal}
\end{equation}
\end{linenomath}
Solving Eq.~(\ref{eq:nodal}) numerically, we obtain the configuration of the nodal lines
as shown in Figs.~\ref{fig:nodal}(a) and \ref{fig:nodal}(b).
The bird's eye view [Fig.~\ref{fig:nodal}(a)] indicates that the nodal lines are 
approximately along the $k_x$ direction, as discussed in Sec.~\ref{sec:firstprinciple},
and are almost located on the planes perpendicular to the $k_z$ axis.
Becuase the crystal has $C_{2y}$ rotation, inversion, and $\sigma_y$ symmetries,
the nodal lines have the same symmetries.
In particular, owing to the $C_{2y}$ rotation symmetry, 
the projection of the nodal lines onto the $k_x$-$k_y$ plane [Fig.~\ref{fig:nodal}(b)]
intersect on the $k_x=0$ line.
Figure~\ref{fig:nodal}(c) shows the Fermi surface of the proposed tight-binding model,
where the purple and green surfaces represent the electron and hole pockets, respectively. 
The energy at the Dirac node point fluctuates along the nodal lines,
which results in alternating thin tube-shaped electron and hole pockets.

\section{CDW phase \label{sec:cdw}}
HMTSF-TCNQ exhibits a CDW phase transition at $T_c\simeq30$ K.
Theoretically, the nesting vector of this CDW state is presumed to be the vector between Dirac points 
on the nodal lines at the Fermi level, as shown in Fig.~\ref{fig:nodal}(a).
In the proposed model, the nesting vector is ${\bm q} \simeq (0,0, 0.937\pi/c \cos{\beta} )$.
Although the obtained nesting vector is slightly different from the experimental value \cite{ravy}, 
${\bm q} \simeq (0,0, 0.74\pi/c \cos{\beta} )$,
we use the former in the following analyses.

Considering the nesting in the mean-field approximation \cite{frorich,kuper,rice-strassler}, the effective model becomes 
\begin{linenomath}
\begin{equation}
	H_{\rm CDW}(\k)=
	\left(
	\begin{array}{cccc}
		t_{\HH}(\k) & t^*_{\HT}(\k) & \Delta & 0 \\
		t_{\HT}(\k) & t_{\TT}(\k) & 0 & \Delta \\
		\Delta & 0 & t_{\HH}(\k-{\bm q}) & t^*_{\HT}(\k-{\bm q}) \\
		0 & \Delta & t_{\HT}(\k-{\bm q}) & t_{\TT}(\k-{\bm q}) \\
	\end{array}
	\right),
	\label{eq:cdwham}
\end{equation}
\end{linenomath}
where $\Delta$ is the order parameter of the CDW state,
which is assumed to be common in the HMTSF and TCNQ chain for simplicity.
Note that the periodicity in the $k_z$ direction is lost and that $k_z$ 
is only allowed in the vicinity of the nodal lines in the $k_z>0$ region ($k_z\simeq  0.38$\AA${}^{-1}$).
However, the nodal lines are located on almost the same plane as in the case of $\Delta=0$; therefore,
we can still discuss the nodal lines using the effective Hamiltonian in Eq.~(\ref{eq:cdwham}).

\begin{figure*}
\begin{center}
\includegraphics[width=17.8cm]{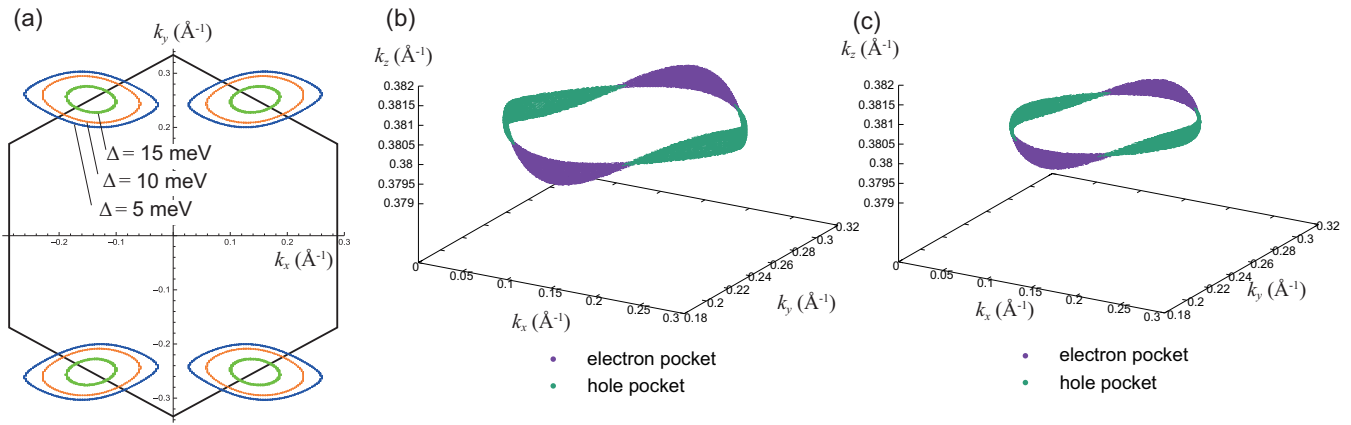}
\caption{
(a) Projection of nodal lines onto the $k_x$-$k_y$ plane for the CDW phase.
The blue, orange, and green lines correspond to $\Delta$=5, 10, and 15 meV, respectively.
Electron and hole pockets for (b) $\Delta=5$ meV and (c) $\Delta=10$ meV.
The purple and green surfaces represent the electron and hole pockets, respectively.}
\label{fig:fermisurfacescdw}
\end{center}
\end{figure*}

Diagonalizing the effective model of Eq.~(\ref{eq:cdwham}), 
we obtain the $\Delta$ dependence of the nodal lines, as shown in Fig.~\ref{fig:fermisurfacescdw}(a).
The blue, orange, and green lines represent the nodal lines with $\Delta=$5, 10, and 15 meV, respectively.
For the case of $\Delta=0$ meV [Fig.~\ref{fig:nodal}(b)], the nodal lines touch each other at a point.
However, with finite $\Delta$, a band gap opens at this point,
which is connected by the nesting vector.
As a result, closed nodal lines are formed.
As $\Delta$ increases, the closed nodal lines become small 
and vanish when $\Delta \simeq18$ meV.
Figures~\ref{fig:fermisurfacescdw}(b) and \ref{fig:fermisurfacescdw}(c) show 
the Fermi surfaces for the cases with $\Delta =5$ meV and $\Delta=10$ meV, respectively.
They consist of alternating electron and hole pockets
along the nodal lines, as in the case of $\Delta =0$ meV,
and these CDW states are still semimetallic.

\section{physical quantities\label{sec:physq}}
In this section, we evaluate two physical quantities that are important 
to the experiments: the spin-lattice relaxation time $T_1$ and 
orbital magnetic susceptibility.
\subsection{Spin-lattice relaxation time $T_1$ }
\begin{figure}
\begin{center}
\includegraphics[width=\linewidth]{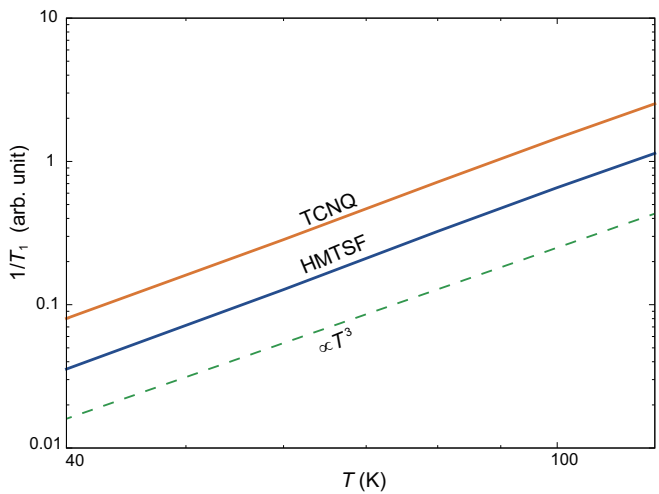}
\caption{Temperature dependence of $1/T_1$ above $T_c$.
The blue and orange lines represent the theoretical values for the HMTSF and TCNQ molecules, respectively.
The powers for the two molecules are approximately $3.1$.
The power $T^{3}$ is also indicated by a dashed line for reference.}
\label{fig:t1}
\end{center}
\end{figure}

The spin-lattice relaxation time $T_1$ reflects the electronic state near the Fermi level.
To experimentally confirm the existence of nodal lines, we theoretically evaluate $T_1$ 
for the normal phase.
Although there are several types of the origins of spin-lattice relaxation, in this analysis,
we focus on the Fermi contact term between the conduction electron and the nucleus in 
each molecule, which is given by 
\cite{moriya,suzumura89,katayama},
\begin{linenomath}
\begin{align}
	\frac{1}{(T_1)_\alpha}\propto T \int_{-\infty}^\infty [D_\alpha(\varepsilon)]^2 
	\left(-f'(\varepsilon)\right) d\varepsilon,
	\label{eq:t1}
\end{align}
\end{linenomath}
where $\alpha$ indicates HMTSF or TCNQ, $D_\alpha(\varepsilon)$ is the DOS for the orbital $\alpha$, and 
$f'(\varepsilon)$ is the derivative of the Fermi distribution function 
$f(\varepsilon)=[1+e^{\beta(\varepsilon-\mu)}]^{-1}$.
We assume that the chemical potential does not change with temperature above $T_c$.
Figure~\ref{fig:t1} shows the temperature dependence of $1/T_1$ in the logscale without the CDW order parameter. 
The blue (orange) solid line represents the theoretical values 
evaluated by Eq.~(\ref{eq:t1}) for the HMTSF (TCNQ) molecule.
The power $T^{3}$ is also shown for reference.
We observe that $(T_1)_\mathrm{HMTSF}$ and $(T_1)_{\rm TCNQ}$ follow the power law $1/T_1\propto T^{3.1}$,
which is very close to the expected law $1/T_1\propto T^{3}$ for pure two-dimensional Dirac electrons.
\subsection{Orbital magnetic susceptibility \label{sec:mag}}
As mentioned in Sec.~\ref{sec:intro}, HMTSF-TCNQ shows a large diamagnetic susceptibility at low temperatures.
To understand the origin and the temperature dependence, 
we evaluate the orbital magnetic susceptibility in the proposed model with nodal lines 
using the Fukuyama formula \cite{fukuyama71} given by,
\begin{linenomath}
\begin{equation}
	\chi_{\rm orbit}=\frac{e^2}{\hbar^2}\sum_{n\k} {\rm Tr} \,\gamma_x \mathcal{G}\gamma_y
	\mathcal{G}\gamma_x\mathcal{G}\gamma_y \mathcal{G},
	\label{eq:fukuyama}
\end{equation}
\end{linenomath}
where $\mathcal{G}$ represents the thermal Green's function
\begin{linenomath}
\begin{equation}
	\mathcal{G}=[i\varepsilon_n - H_{\rm CDW}(\k) + \mu + i\Gamma {\rm sign}({\varepsilon_n})]^{-1},
\end{equation}
\end{linenomath}
where $\varepsilon_n=(2n+1)\pi k_{\rm B} T$, $\Gamma$, and $\mu$ are the Matsubara frequency, 
the damping rate of the electron,
and the chemical potential, respectively.
$\gamma_i$ is the current operator in the $i(=x\, {\rm or}\, y)$ direction,
which is defined by 
$\gamma_i=\partial H_{\rm CDW}(\k)/\partial k_i$, and
the $n$ summation means the sum over the Matsubara frequency $\varepsilon_n$.
The chemical potential is determined by the charge-neutrality condition, 
\begin{linenomath}
\begin{align}
\sum_{l{\bm k}}f(\varepsilon_{l{\bm k}}) =\sum_{l{\bm k}}[1-f(\varepsilon_{l{\bm k}})], \label{eq:fermidist}
\end{align}
\end{linenomath}
where $\varepsilon_{l\k}$ is the energy dispersion of the $l$th band.

For simplicity, we assume the temperature dependence of the order parameter,
as typically used for the Bardeen-Cooper-Schrieffer superconductivity \cite{fetterwalecka},
\begin{linenomath}
\begin{equation}
	\Delta(T)=\Delta_0 \sqrt{1-\frac{T}{T_c}},\label{eq:delta}
\end{equation}
\end{linenomath}
where $\Delta_0$ is the order parameter at zero temperature.
$T_c$ is set to 30 K, which is in accordance with the experiment \cite{cooper76}.
In the following analysis, we evaluate the chemical potential by only considering 
the temperature dependence through Eq.~(\ref{eq:delta}) and 
setting $\beta\to \infty$ in Eq.~(\ref{eq:fermidist}).
This approximation is justified when $T\simeq 0$; however,
we expect that it is still valid at high temperatures 
because the thermal fluctuation overpowers the error of this approximation.

\begin{figure}
\begin{center}
\includegraphics[width=\linewidth]{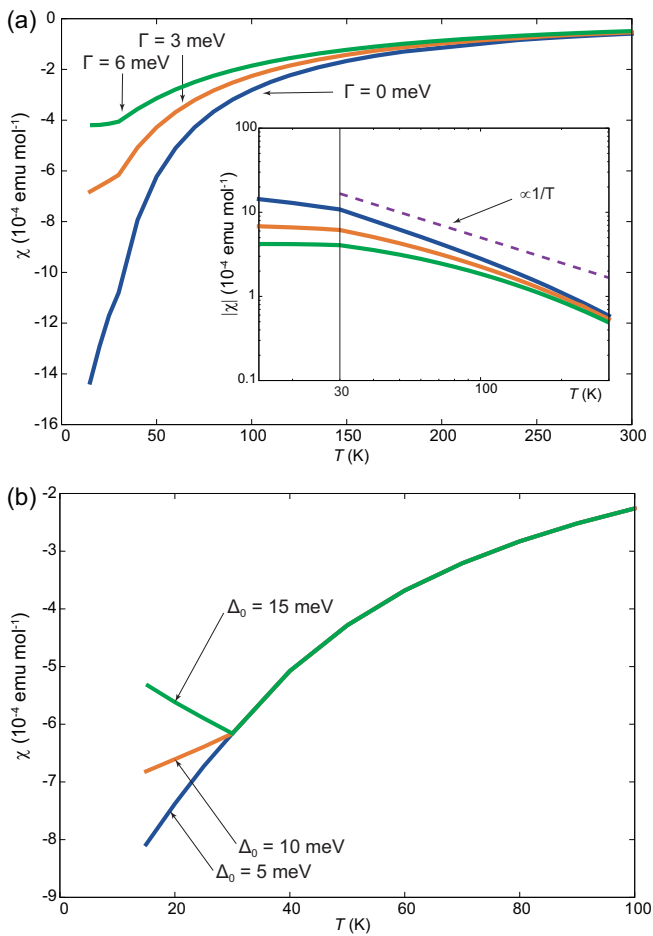}
\caption{(a) Orbital magnetic susceptibility for $\Delta_0=10$ meV
as a function of the temperature.
The blue, orange, and green lines correspond to $\Gamma=0$ meV, 3 meV, and 6 meV, respectively.
Inset: Absolute value of the orbital magnetic susceptibility in the logscale.
The power $T^{-1}$ is also shown for reference.
(b) Orbital magnetic susceptibility for $\Gamma=3$ meV
as a function of the temperature.
The blue, orange, and green lines correspond to $\Delta_0$=5 meV, 10 meV, 
and 15 meV, respectively.}
\label{fig:chi-th}
\end{center}
\end{figure}

Integrating Eq.~(\ref{eq:fukuyama}) numerically, 
we obtain the temperature dependence of the orbital magnetic susceptibility
as shown in Fig.~\ref{fig:chi-th}(a) for $\Delta_0$=10 meV ($\Gamma=0, 3, \textrm{and}\, 6$ meV).
Above $T_c$, we observe that the orbital magnetic susceptibility increases negatively 
as the temperature decreases from the room temperature for every $\Gamma$.  
The nodal lines are the ensembles of Dirac electrons; therefore
the Landau-Peierls contribution (the extension of the Landau diamagnetism 
to periodic systems) is expected to be very small \cite{raoux-piechon,ogata3}, and 
the diamagnetism is attributed to the interband effect \cite{fukuyama-kubo,fukuyama07,fuseya,raoux-piechon}. 
Interband diamagnetism in two-dimensional Dirac electrons has a temperature dependence of $T^{-1}$
in the clean limit \cite{mcclure}.
The inset of Fig.~\ref{fig:chi-th}(a) indicates that the present numerical result 
for the $\Gamma=0$ meV case (shown in blue)
asymptotically obeys the power law for the low temperatures ($\sim$30 K).
For the region $T_c < T< 100$ K, impurities affect the interband effect.
Although the orbital magnetism in \textit{massive} Dirac electron systems 
is not significantly sensitive to impurities \cite{fuseya},
that in \textit{massless} Dirac electrons is rather sensitive \cite{fukuyama07,nakamura07}.

Below $T_c$, we experience the effect of the nodal line deformation as well as the interband effect
and impurity scattering.
The inflection at $T_c$ is due to the order parameter.
As the temperature decreases,
the order parameter increases, and the nodal lines shrink, as discussed in the previous section.
This implies a decrease in the number of the Dirac electrons,
which results in suppression of diamagnetism.
The relation between the magnitude of diamagnetism and the size of 
the nodal lines is discussed in detail in Ref.~\cite{tateishi}.
Note that a gap does not open in the DOS as long as $\Delta_0<18$ meV,
and \textit{massless} Dirac electrons are present for the cases shown in Fig.~\ref{fig:chi-th}(a).
Therefore, the diamagnetism diverges at zero temperature for the case of $\Gamma=0$ meV,
whereas the divergence is suppressed with a finite $\Gamma$. 

Figure~\ref{fig:chi-th}(b) shows the orbital magnetic susceptibility for the cases 
of $\Delta_0=5,10,\textrm{and}\,15$ meV, where we set $\Gamma=3$ meV as a typical value.
For the case with a small $\Delta_0$, the diamagnetism monotonically increases as the temperature decreases.
However, for the case with large $\Delta_0$, the diamagnetism is suppressed because
the number of Dirac electrons decreases rapidly.
Whether $\chi_{\rm orbit}$ decreases or increases as the temperature decreases depends on the following three effects:
the interband effect, impurity scattering, and nodal line deformation.

\begin{figure}
\begin{center}
\includegraphics[width=\linewidth]{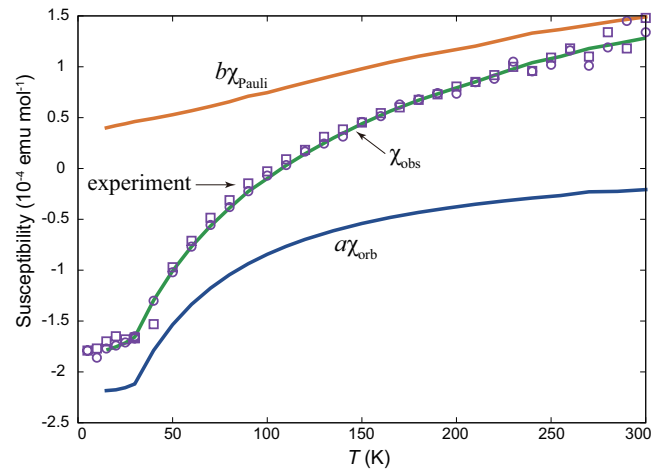}
\caption{
Magnetic susceptibility as a function of temperature.
The blue, orange, and green lines represent the theoretical evaluation of
$a\chi_{\rm orbit}$, $b\chi_{\rm Pauli}$, and $\chi_{\rm obs}$, respectively, 
and the circle (square) markers represent the experimental data obtained by the SQUID magnetometer 
in the cooling (heating) process.}
\label{fig:comparation}
\end{center}
\end{figure}

We compare the above theoretical results with a SQUID magnetometer measurement.
The magnetizations were measured between 2--300 K using a SQUID magnetometer operated by 
a magnetic properties measurement system (Quantum Design Inc., CA, USA).
A newly synthesized polycrystalline sample of $\sim 5$ mg is used.
The core contribution of the diamagnetism estimated by Pascal's law ($-2.8 \times 10^{-4}$ emu/mole) is subtracted,
as in the previous study \cite{soda}.
The experimental data obtained in the cooling (heating) process are shown 
in Fig.~\ref{fig:comparation} with circle (square) markers;
the data reproduce the temperature dependence
obtained in a previous study \cite{soda}, including the absolute values.

Above 100 K, the total magnetic susceptibility is paramagnetic.
However, as the temperature decreases, the diamagnetism prevails,
and finally reaches a plateau below the CDW phase transition temperature, $T_c=30$ K.

To understand the experimental results, we consider that the total magnetic susceptibility
is given by
\begin{linenomath}
\begin{eqnarray}
\chi_{\rm obs} = a\chi_{\rm orbit} + b\chi_{\rm Pauli},
\end{eqnarray}
\end{linenomath}
with positive coefficients $a$ and $b$ and $\chi_{\rm Pauli}$ is the Pauli paramagnetic contribution
\begin{linenomath}
\begin{equation}
	\chi_{\rm Pauli}=-\mu_{\rm B}^2 k_{\rm B}T \sum_{n\k} {\rm Tr} \,\mathcal{G}^2,
\end{equation}
\end{linenomath}
where $\mu_{\rm B}$ is the Bohr magneton $\mu_{\rm B}=|e|\hbar/2m$.
The coefficient $a$ reflects the randomness of the polycrystal orientation,
and it is expected to be smaller than 1
because $\chi_{\rm orbit}$ is calculated under a magnetic field in the direction 
of the largest diamagnetism.
However, the coefficient $b$ ideally equals 1.
Note that $\chi_{\rm orbit}$ ($\chi_{\rm Pauli}$) is dominant at low (high) temperatures.
Thus, the behavior below $T_c$ is primarily determined by $\chi_{\rm orbit}$,
which depends on $\Delta_0$ and $\Gamma$ up to a numerical factor.
To reproduce the plateau that was experimentally observe, we choose $\Delta_0=11$ meV and $\Gamma=4$ meV,
and we set the numerical factor as $a=0.4$.
However, for the $T>100$~K region, the dominant contribution is from $b\chi_{\rm Pauli}$,
and to fit the data, we set $b=2.1$.
We also show the obtained $a\chi_{\rm orbit}$, $b\chi_{\rm Pauli}$, and $\chi_{\rm obs}$ 
in Fig.~\ref{fig:comparation} with solid lines.
The proposed theory is in excellent agreement with the experimental results for all temperature regions.
This suggests that the experimental result is explained by the above scenario,
i.e., the anomalous large diamagnetism at low temperatures is explained by
the conflict between the interband effect, impurity scattering, and the deformation of nodal lines.

It should be noted that the coefficient $b$ deviates from unity in this fitting.
The origin of this deviation is not clear, however, it may be the intramolecular diamagnetism.
Although this contribution has been partly subtracted as Pascal's law,
a further consideration will be needed for a precise evaluation.


\section{Conclusion\label{sec:summary}}
The present study is based on a first-principles calculation and tight-binding analysis,
and it predicts that an organic complex
HMTSF-TCNQ is a new candidate material for the nodal line semimetal.
We have also clarified that the CDW deforms open nodal lines into closed ones.
We evaluated the spin-lattice relaxation time $T_1$ and the magnetic susceptibility.
An experiment was also conducted to investigate the magnetization, and the large anomalous 
diamagnetism, the origin of which had long been controversial, was reproduced.
The present theoretical evaluation of magnetic susceptibility is in excellent agreement with the experimental results,
and it has clarified that the conflict between the interband diamagnetism, impurity scattering, and 
nodal line deformation realizes this anomalous diamagnetism.
Our evaluation of $T_1$ supports the existence of nodal lines, 
and it will be experimentally confirmed by means of 
a nuclear magnetic resonance measurement. 
This study is in progress, and the results will be presented elsewhere.

For this material, many intriguing physical properties besides the magnetic susceptibility
have been investigated,
such as the Seebeck coefficient and Hall conductivity.
These properties are also expected to be derived from the interplay between the CDW and nodal lines.
These characteristics will be elaborated by the framework proposed in this study.

\begin{acknowledgments}
We thank D.~Miyafuji, T.~Mizoguchi, M.~Udagawa, H.~Maebashi, and H.~Fukuyama for fruitful discussions.
We also thank R.\ Sugiura, T.\ Nakamura, and T.\ Takahashi for their magnetic susceptibility measurements,
R. Kato for providing high quality samples, 
and H.~Cui for providing crystal data.
This work was supported by Grant-in-Aid for Scientific Research from 
the Japan Society for the Promotion of Science (Grand No.~JP18H01162, JP19K03720, JP18K03482).
S.O. and I.T. were supported by the Japan Society for Science Promotion (JSPS) 
through the Program for Leading Graduate Schools (MERIT).
\end{acknowledgments}

\appendix
\section{Wave functions and inversion parities at TRIM
\label{sec:wavefunctions}}
\begin{figure*}[h]
\begin{center}
\includegraphics[width=\linewidth]{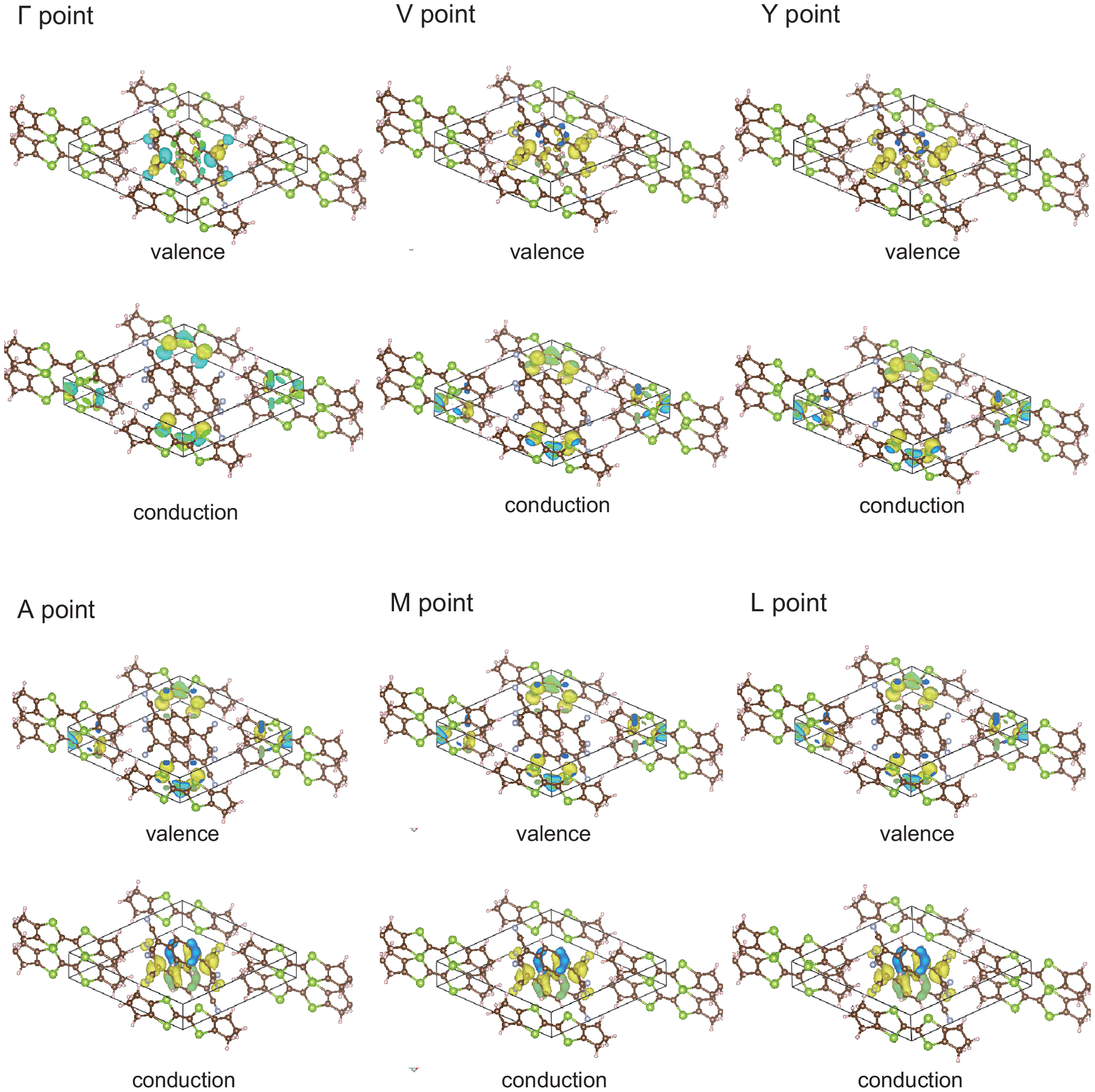}
\caption{Wave functions for the valence and conduction bands at the TRIM.}
\label{fig:wfall}
\end{center}
\end{figure*}

Figure~\ref{fig:wfall} shows the wave functions for the valence and 
conduction bands at the TRIM,
which are obtained by the first-principles calculation.
For the $\Gamma$ point, the colors in the figure represent the signs of the wave functions.
For other $\bm k$ points, the wave functions are generally complex and are shown in yellow.
The cross-sections of the wave functions with the Brillouin zone boundary are shown in blue.
Each wave function is the molecular orbital of either HMTSF or TCNQ.
This suggests that the wave function at a generic point is also
given by the linear combination of the wave functions for the two molecules,
which justifies the assumption that the low-energy electronic state is described 
in terms of these two orbitals.

The parities of the inversion operation for the wave functions at TRIM 
are summarized in Table \ref{table:parity}.
The symmetry indicator is calculated from these values for the valence band.

\begin{table}[h]
	\caption{Parities of inversion operation for the wave functions at TRIM.}
   \label{table:parity}
	 \begin{ruledtabular}
		\begin{tabular}{lcccccc}
  		 & $\Gamma$ & V& Y&  A&M&L \\
			\colrule
  		Valence & $+$ & $-$ & $+$ & $-$ & $+$ & $+$  \\
			Conduction & $-$ & $-$ & $-$ & $-$ & $+$ & $-$ \\
  	\end{tabular}
	\end{ruledtabular}
\end{table}

\section{Dirac cone\label{sec:diraccone}}
\begin{figure}[h]
\begin{center}
\includegraphics[width=\linewidth]{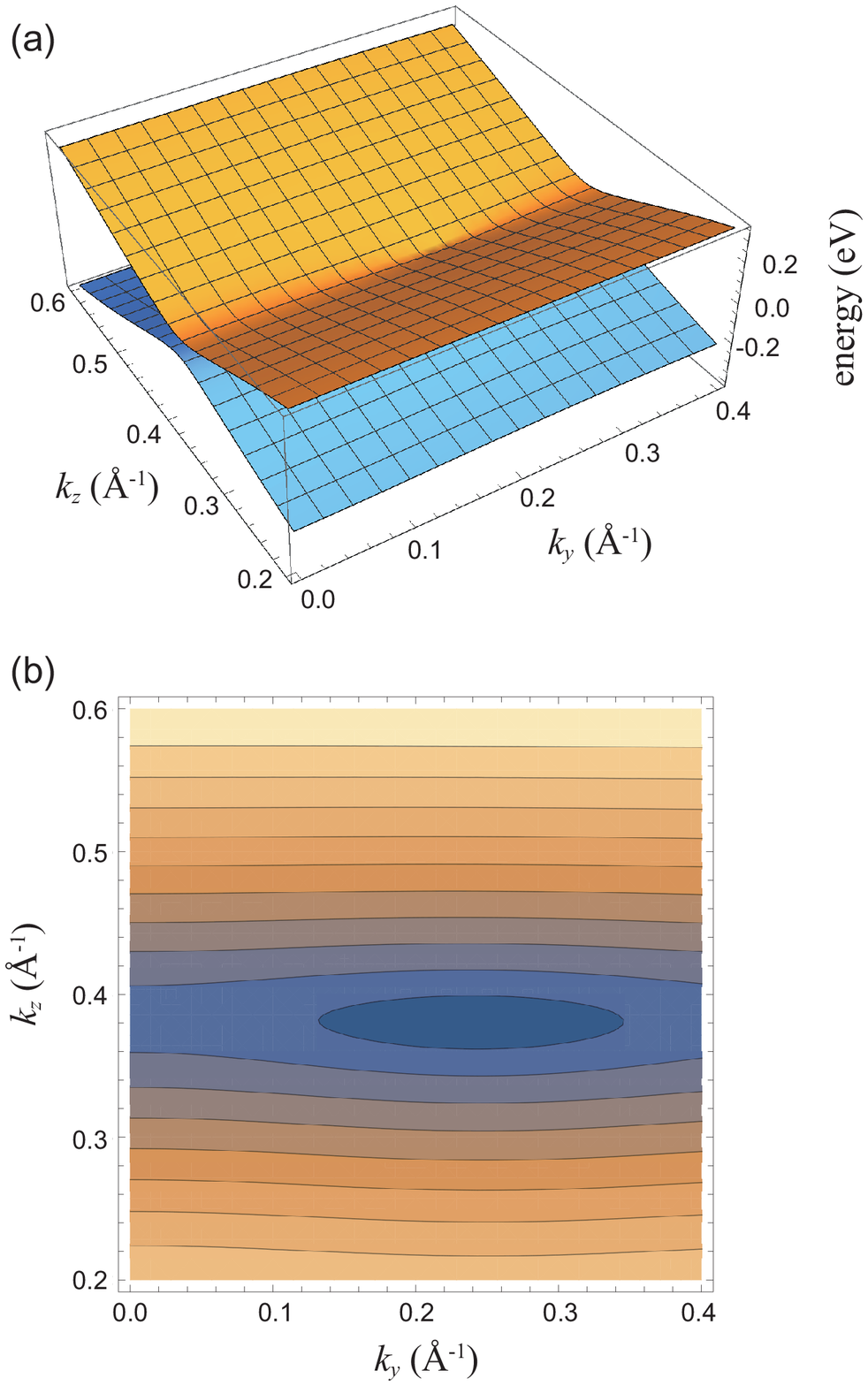}
\caption{(a) Energy dispersion near a Dirac point when $k_x=0$.
The orange (blue) surface corresponds to $E_+$ ($E_-$).
(b) Contour plot of $E_+-E_-$.}
\label{fig:diraccone}
\end{center}
\end{figure}

	The energy dispersion near a Dirac point when $k_x=0$ is shown in Fig.~\ref{fig:diraccone}(a),
and the contour plot of the difference of the energies $E_+-E_-$ is shown in 
Fig.~\ref{fig:diraccone}(b).
We observed that the Dirac cone has strong anisotropy, and the system is two-dimensional 
only in the vicinity of the Dirac point.

\bibliography{hmtsftcnq}

\providecommand{\noopsort}[1]{}\providecommand{\singleletter}[1]{#1}%
\begin{thebibliography}{55}%
\makeatletter
\providecommand \@ifxundefined [1]{%
 \@ifx{#1\undefined}
}%
\providecommand \@ifnum [1]{%
 \ifnum #1\expandafter \@firstoftwo
 \else \expandafter \@secondoftwo
 \fi
}%
\providecommand \@ifx [1]{%
 \ifx #1\expandafter \@firstoftwo
 \else \expandafter \@secondoftwo
 \fi
}%
\providecommand \natexlab [1]{#1}%
\providecommand \enquote  [1]{``#1''}%
\providecommand \bibnamefont  [1]{#1}%
\providecommand \bibfnamefont [1]{#1}%
\providecommand \citenamefont [1]{#1}%
\providecommand \href@noop [0]{\@secondoftwo}%
\providecommand \href [0]{\begingroup \@sanitize@url \@href}%
\providecommand \@href[1]{\@@startlink{#1}\@@href}%
\providecommand \@@href[1]{\endgroup#1\@@endlink}%
\providecommand \@sanitize@url [0]{\catcode `\\12\catcode `\$12\catcode
  `\&12\catcode `\#12\catcode `\^12\catcode `\_12\catcode `\%12\relax}%
\providecommand \@@startlink[1]{}%
\providecommand \@@endlink[0]{}%
\providecommand \url  [0]{\begingroup\@sanitize@url \@url }%
\providecommand \@url [1]{\endgroup\@href {#1}{\urlprefix }}%
\providecommand \urlprefix  [0]{URL }%
\providecommand \Eprint [0]{\href }%
\providecommand \doibase [0]{https://doi.org/}%
\providecommand \selectlanguage [0]{\@gobble}%
\providecommand \bibinfo  [0]{\@secondoftwo}%
\providecommand \bibfield  [0]{\@secondoftwo}%
\providecommand \translation [1]{[#1]}%
\providecommand \BibitemOpen [0]{}%
\providecommand \bibitemStop [0]{}%
\providecommand \bibitemNoStop [0]{.\EOS\space}%
\providecommand \EOS [0]{\spacefactor3000\relax}%
\providecommand \BibitemShut  [1]{\csname bibitem#1\endcsname}%
\let\auto@bib@innerbib\@empty
\bibitem [{\citenamefont {Qin}\ \emph {et~al.}(2020)\citenamefont {Qin},
  \citenamefont {Li}, \citenamefont {Du}, \citenamefont {Wang}, \citenamefont
  {Zhang}, \citenamefont {Yu}, \citenamefont {Lu},\ and\ \citenamefont
  {Xie}}]{3dhall}%
  \BibitemOpen
  \bibfield  {author} {\bibinfo {author} {\bibfnamefont {F.}~\bibnamefont
  {Qin}}, \bibinfo {author} {\bibfnamefont {S.}~\bibnamefont {Li}}, \bibinfo
  {author} {\bibfnamefont {Z.~Z.}\ \bibnamefont {Du}}, \bibinfo {author}
  {\bibfnamefont {C.~M.}\ \bibnamefont {Wang}}, \bibinfo {author}
  {\bibfnamefont {W.}~\bibnamefont {Zhang}}, \bibinfo {author} {\bibfnamefont
  {D.}~\bibnamefont {Yu}}, \bibinfo {author} {\bibfnamefont {H.-Z.}\
  \bibnamefont {Lu}},\ and\ \bibinfo {author} {\bibfnamefont {X.~C.}\
  \bibnamefont {Xie}},\ }\bibfield  {title} {\bibinfo {title} {Theory for the
  charge-density-wave mechanism of 3d quantum {Hall} effect},\ }\href
  {https://doi.org/10.1103/PhysRevLett.125.206601} {\bibfield  {journal}
  {\bibinfo  {journal} {Phys. Rev. Lett.}\ }\textbf {\bibinfo {volume} {125}},\
  \bibinfo {pages} {206601} (\bibinfo {year} {2020})}\BibitemShut {NoStop}%
\bibitem [{\citenamefont {Gooth}\ \emph {et~al.}(2019)\citenamefont {Gooth},
  \citenamefont {Bradlyn}, \citenamefont {Honnali}, \citenamefont {Schindler},
  \citenamefont {Kumar}, \citenamefont {Noky}, \citenamefont {Qi},
  \citenamefont {Shekhar}, \citenamefont {Sun}, \citenamefont {Wang},
  \citenamefont {Bernevig},\ and\ \citenamefont {Fleser}}]{axionic}%
  \BibitemOpen
  \bibfield  {author} {\bibinfo {author} {\bibfnamefont {J.}~\bibnamefont
  {Gooth}}, \bibinfo {author} {\bibfnamefont {B.}~\bibnamefont {Bradlyn}},
  \bibinfo {author} {\bibfnamefont {S.}~\bibnamefont {Honnali}}, \bibinfo
  {author} {\bibfnamefont {C.}~\bibnamefont {Schindler}}, \bibinfo {author}
  {\bibfnamefont {N.}~\bibnamefont {Kumar}}, \bibinfo {author} {\bibfnamefont
  {J.}~\bibnamefont {Noky}}, \bibinfo {author} {\bibfnamefont {Y.}~\bibnamefont
  {Qi}}, \bibinfo {author} {\bibfnamefont {C.}~\bibnamefont {Shekhar}},
  \bibinfo {author} {\bibfnamefont {Y.}~\bibnamefont {Sun}}, \bibinfo {author}
  {\bibfnamefont {Z.}~\bibnamefont {Wang}}, \bibinfo {author} {\bibfnamefont
  {B.~A.}\ \bibnamefont {Bernevig}},\ and\ \bibinfo {author} {\bibfnamefont
  {C.}~\bibnamefont {Fleser}},\ }\bibfield  {title} {\bibinfo {title} {Axionic
  charge-density wave in the {Weyl} semimetal ({TaSe}$_4$)$_2${I}},\
  }\href@noop {} {\bibfield  {journal} {\bibinfo  {journal} {Nature}\ }\textbf
  {\bibinfo {volume} {575}},\ \bibinfo {pages} {315} (\bibinfo {year}
  {2019})}\BibitemShut {NoStop}%
\bibitem [{\citenamefont {Kobayashi}\ \emph {et~al.}(2007)\citenamefont
  {Kobayashi}, \citenamefont {Katayama}, \citenamefont {Suzumura},\ and\
  \citenamefont {Fukuyama}}]{kobayashi07}%
  \BibitemOpen
  \bibfield  {author} {\bibinfo {author} {\bibfnamefont {A.}~\bibnamefont
  {Kobayashi}}, \bibinfo {author} {\bibfnamefont {S.}~\bibnamefont {Katayama}},
  \bibinfo {author} {\bibfnamefont {Y.}~\bibnamefont {Suzumura}},\ and\
  \bibinfo {author} {\bibfnamefont {H.}~\bibnamefont {Fukuyama}},\ }\bibfield
  {title} {\bibinfo {title} {Massless fermions in organic conductor},\
  }\href@noop {} {\bibfield  {journal} {\bibinfo  {journal} {J. Phys. Soc.
  Jpn.}\ }\textbf {\bibinfo {volume} {76}},\ \bibinfo {pages} {034711}
  (\bibinfo {year} {2007})}\BibitemShut {NoStop}%
\bibitem [{\citenamefont {Kobayashi}\ \emph {et~al.}(2008)\citenamefont
  {Kobayashi}, \citenamefont {Suzumura},\ and\ \citenamefont
  {Fukuyama}}]{kobayashi08}%
  \BibitemOpen
  \bibfield  {author} {\bibinfo {author} {\bibfnamefont {A.}~\bibnamefont
  {Kobayashi}}, \bibinfo {author} {\bibfnamefont {Y.}~\bibnamefont
  {Suzumura}},\ and\ \bibinfo {author} {\bibfnamefont {H.}~\bibnamefont
  {Fukuyama}},\ }\bibfield  {title} {\bibinfo {title} {Hall effect and orbital
  diamagnetism in zerogap state of molecular conductor
  $\alpha$-({BEDT-TTF})${}_2${I}${}_3$},\ }\href@noop {} {\bibfield  {journal}
  {\bibinfo  {journal} {J. Phys. Soc. Jpn.}\ }\textbf {\bibinfo {volume}
  {77}},\ \bibinfo {pages} {064718} (\bibinfo {year} {2008})}\BibitemShut
  {NoStop}%
\bibitem [{\citenamefont {Wang}\ \emph {et~al.}(2013)\citenamefont {Wang},
  \citenamefont {Liu},\ and\ \citenamefont {Liu}}]{wang-prl}%
  \BibitemOpen
  \bibfield  {author} {\bibinfo {author} {\bibfnamefont {Z.~F.}\ \bibnamefont
  {Wang}}, \bibinfo {author} {\bibfnamefont {Z.}~\bibnamefont {Liu}},\ and\
  \bibinfo {author} {\bibfnamefont {F.}~\bibnamefont {Liu}},\ }\bibfield
  {title} {\bibinfo {title} {Quantum anomalous {Hall} effect in {2D} organic
  topological insulators},\ }\href
  {https://doi.org/10.1103/PhysRevLett.110.196801} {\bibfield  {journal}
  {\bibinfo  {journal} {Phys. Rev. Lett.}\ }\textbf {\bibinfo {volume} {110}},\
  \bibinfo {pages} {196801} (\bibinfo {year} {2013})}\BibitemShut {NoStop}%
\bibitem [{\citenamefont {Zhang}\ \emph {et~al.}(2016)\citenamefont {Zhang},
  \citenamefont {Wang}, \citenamefont {Huang}, \citenamefont {Cui},
  \citenamefont {Wang}, \citenamefont {Du}, \citenamefont {Gao},\ and\
  \citenamefont {Liu}}]{zhang-nanoletter}%
  \BibitemOpen
  \bibfield  {author} {\bibinfo {author} {\bibfnamefont {L.~Z.}\ \bibnamefont
  {Zhang}}, \bibinfo {author} {\bibfnamefont {Z.~F.}\ \bibnamefont {Wang}},
  \bibinfo {author} {\bibfnamefont {B.}~\bibnamefont {Huang}}, \bibinfo
  {author} {\bibfnamefont {B.}~\bibnamefont {Cui}}, \bibinfo {author}
  {\bibfnamefont {Z.}~\bibnamefont {Wang}}, \bibinfo {author} {\bibfnamefont
  {S.~X.}\ \bibnamefont {Du}}, \bibinfo {author} {\bibfnamefont {H.-J.}\
  \bibnamefont {Gao}},\ and\ \bibinfo {author} {\bibfnamefont {F.}~\bibnamefont
  {Liu}},\ }\bibfield  {title} {\bibinfo {title} {Intrinsic two-dimensional
  organic topological insulators in metal-dicyanoanthracene lattices},\
  }\href@noop {} {\bibfield  {journal} {\bibinfo  {journal} {Nano Letters}\
  }\textbf {\bibinfo {volume} {16}},\ \bibinfo {pages} {2072} (\bibinfo {year}
  {2016})}\BibitemShut {NoStop}%
\bibitem [{\citenamefont {Liu}\ \emph {et~al.}(2018)\citenamefont {Liu},
  \citenamefont {Wang}, \citenamefont {Wang}, \citenamefont {Yang},\ and\
  \citenamefont {Liu}}]{liu-prb}%
  \BibitemOpen
  \bibfield  {author} {\bibinfo {author} {\bibfnamefont {Z.}~\bibnamefont
  {Liu}}, \bibinfo {author} {\bibfnamefont {H.}~\bibnamefont {Wang}}, \bibinfo
  {author} {\bibfnamefont {Z.~F.}\ \bibnamefont {Wang}}, \bibinfo {author}
  {\bibfnamefont {J.}~\bibnamefont {Yang}},\ and\ \bibinfo {author}
  {\bibfnamefont {F.}~\bibnamefont {Liu}},\ }\bibfield  {title} {\bibinfo
  {title} {Pressure-induced organic topological nodal-line semimetal in the
  three-dimensional molecular crystal $\mathrm{Pd}{(\mathrm{dddt})}_{2}$},\
  }\href {https://doi.org/10.1103/PhysRevB.97.155138} {\bibfield  {journal}
  {\bibinfo  {journal} {Phys. Rev. B}\ }\textbf {\bibinfo {volume} {97}},\
  \bibinfo {pages} {155138} (\bibinfo {year} {2018})}\BibitemShut {NoStop}%
\bibitem [{\citenamefont {Kato}\ \emph {et~al.}(2017)\citenamefont {Kato},
  \citenamefont {Cui}, \citenamefont {Tsumuraya}, \citenamefont {Miyazaki},\
  and\ \citenamefont {Suzumura}}]{kato-jacs}%
  \BibitemOpen
  \bibfield  {author} {\bibinfo {author} {\bibfnamefont {R.}~\bibnamefont
  {Kato}}, \bibinfo {author} {\bibfnamefont {H.}~\bibnamefont {Cui}}, \bibinfo
  {author} {\bibfnamefont {T.}~\bibnamefont {Tsumuraya}}, \bibinfo {author}
  {\bibfnamefont {T.}~\bibnamefont {Miyazaki}},\ and\ \bibinfo {author}
  {\bibfnamefont {Y.}~\bibnamefont {Suzumura}},\ }\bibfield  {title} {\bibinfo
  {title} {Emergence of the {Dirac} electron system in a single-component
  molecular conductor under high pressure},\ }\href@noop {} {\bibfield
  {journal} {\bibinfo  {journal} {J. Am. Chem. Soc.}\ }\textbf {\bibinfo
  {volume} {139}},\ \bibinfo {pages} {1770} (\bibinfo {year}
  {2017})}\BibitemShut {NoStop}%
\bibitem [{\citenamefont {Kato}\ and\ \citenamefont
  {Suzumura}(2020)}]{kato-jpsj}%
  \BibitemOpen
  \bibfield  {author} {\bibinfo {author} {\bibfnamefont {R.}~\bibnamefont
  {Kato}}\ and\ \bibinfo {author} {\bibfnamefont {Y.}~\bibnamefont
  {Suzumura}},\ }\bibfield  {title} {\bibinfo {title} {A tight-binding model of
  an ambient-pressure molecular {Dirac} electron system with open nodal
  lines},\ }\href@noop {} {\bibfield  {journal} {\bibinfo  {journal} {J. Phys.
  Soc. Jpn.}\ }\textbf {\bibinfo {volume} {89}},\ \bibinfo {pages} {044713}
  (\bibinfo {year} {2020})}\BibitemShut {NoStop}%
\bibitem [{\citenamefont {Kawamura}\ \emph {et~al.}(2020)\citenamefont
  {Kawamura}, \citenamefont {Ohki}, \citenamefont {Zhou}, \citenamefont
  {Kobayashi},\ and\ \citenamefont {Kobayashi}}]{kawamura-jpsj}%
  \BibitemOpen
  \bibfield  {author} {\bibinfo {author} {\bibfnamefont {T.}~\bibnamefont
  {Kawamura}}, \bibinfo {author} {\bibfnamefont {D.}~\bibnamefont {Ohki}},
  \bibinfo {author} {\bibfnamefont {B.}~\bibnamefont {Zhou}}, \bibinfo {author}
  {\bibfnamefont {A.}~\bibnamefont {Kobayashi}},\ and\ \bibinfo {author}
  {\bibfnamefont {A.}~\bibnamefont {Kobayashi}},\ }\bibfield  {title} {\bibinfo
  {title} {Tight-binding model and electronic property of {Dirac} nodal line in
  single-component molecular conductor [pt(dmdt)${}_2$]},\ }\href@noop {}
  {\bibfield  {journal} {\bibinfo  {journal} {J. Phys. Soc. Jpn.}\ }\textbf
  {\bibinfo {volume} {89}},\ \bibinfo {pages} {074704} (\bibinfo {year}
  {2020})}\BibitemShut {NoStop}%
\bibitem [{\citenamefont {Kato}\ \emph {et~al.}(2020)\citenamefont {Kato},
  \citenamefont {Cui}, \citenamefont {Minamidate}, \citenamefont {Yeung},\ and\
  \citenamefont {Suzumura}}]{kato2020}%
  \BibitemOpen
  \bibfield  {author} {\bibinfo {author} {\bibfnamefont {R.}~\bibnamefont
  {Kato}}, \bibinfo {author} {\bibfnamefont {H.}~\bibnamefont {Cui}}, \bibinfo
  {author} {\bibfnamefont {T.}~\bibnamefont {Minamidate}}, \bibinfo {author}
  {\bibfnamefont {H.~H.-M.}\ \bibnamefont {Yeung}},\ and\ \bibinfo {author}
  {\bibfnamefont {Y.}~\bibnamefont {Suzumura}},\ }\bibfield  {title} {\bibinfo
  {title} {Electronic structure of a single-component molecular conductor
  [{Pd}(dddt)2] (dddt = 5,6-dihydro-1,4-dithiin-2,3-dithiolate) under high
  pressure},\ }\href@noop {} {\bibfield  {journal} {\bibinfo  {journal} {J.
  Phys. Soc. Jpn.}\ }\textbf {\bibinfo {volume} {89}},\ \bibinfo {pages}
  {124706} (\bibinfo {year} {2020})}\BibitemShut {NoStop}%
\bibitem [{\citenamefont {Bloch}\ \emph {et~al.}(1975)\citenamefont {Bloch},
  \citenamefont {Cowan}, \citenamefont {Bechgaard}, \citenamefont {Pyle},
  \citenamefont {Banks},\ and\ \citenamefont {Poehler}}]{bloch75}%
  \BibitemOpen
  \bibfield  {author} {\bibinfo {author} {\bibfnamefont {A.~N.}\ \bibnamefont
  {Bloch}}, \bibinfo {author} {\bibfnamefont {D.~O.}\ \bibnamefont {Cowan}},
  \bibinfo {author} {\bibfnamefont {K.}~\bibnamefont {Bechgaard}}, \bibinfo
  {author} {\bibfnamefont {R.~E.}\ \bibnamefont {Pyle}}, \bibinfo {author}
  {\bibfnamefont {R.~H.}\ \bibnamefont {Banks}},\ and\ \bibinfo {author}
  {\bibfnamefont {T.~O.}\ \bibnamefont {Poehler}},\ }\bibfield  {title}
  {\bibinfo {title} {Low-temperature metallic behavior and resistance minimum
  in a new quasi one-dimensional organic conductor},\ }\href
  {https://doi.org/10.1103/PhysRevLett.34.1561} {\bibfield  {journal} {\bibinfo
   {journal} {Phys. Rev. Lett.}\ }\textbf {\bibinfo {volume} {34}},\ \bibinfo
  {pages} {1561} (\bibinfo {year} {1975})}\BibitemShut {NoStop}%
\bibitem [{\citenamefont {Bechgaard}\ \emph {et~al.}(1976)\citenamefont
  {Bechgaard}, \citenamefont {Cowan},\ and\ \citenamefont {Bloch}}]{bechgaard}%
  \BibitemOpen
  \bibfield  {author} {\bibinfo {author} {\bibfnamefont {K.}~\bibnamefont
  {Bechgaard}}, \bibinfo {author} {\bibfnamefont {D.~O.}\ \bibnamefont
  {Cowan}},\ and\ \bibinfo {author} {\bibfnamefont {A.~N.}\ \bibnamefont
  {Bloch}},\ }\bibfield  {title} {\bibinfo {title} {Stabilization of the
  organic metallic state: The properties of two substituted
  tetraselenafulvalenes and their {TCNQ} salts},\ }\href@noop {} {\bibfield
  {journal} {\bibinfo  {journal} {Molecular Crystals and Liquid Crystals}\
  }\textbf {\bibinfo {volume} {32}},\ \bibinfo {pages} {227} (\bibinfo {year}
  {1976})}\BibitemShut {NoStop}%
\bibitem [{\citenamefont {Soda}\ \emph {et~al.}(1976)\citenamefont {Soda},
  \citenamefont {J\`{e}rome}, \citenamefont {Weger}, \citenamefont
  {Bechgaard},\ and\ \citenamefont {Pedersen}}]{soda}%
  \BibitemOpen
  \bibfield  {author} {\bibinfo {author} {\bibfnamefont {G.}~\bibnamefont
  {Soda}}, \bibinfo {author} {\bibfnamefont {D.}~\bibnamefont {J\`{e}rome}},
  \bibinfo {author} {\bibfnamefont {M.}~\bibnamefont {Weger}}, \bibinfo
  {author} {\bibfnamefont {K.}~\bibnamefont {Bechgaard}},\ and\ \bibinfo
  {author} {\bibfnamefont {E.}~\bibnamefont {Pedersen}},\ }\bibfield  {title}
  {\bibinfo {title} {Spin relaxation and magnetic susceptibility studies of
  {HMTSF-TCNQ}},\ }\href@noop {} {\bibfield  {journal} {\bibinfo  {journal}
  {Solid State Commun.}\ }\textbf {\bibinfo {volume} {20}},\ \bibinfo {pages}
  {107} (\bibinfo {year} {1976})}\BibitemShut {NoStop}%
\bibitem [{\citenamefont {Weger}(1976)}]{weger}%
  \BibitemOpen
  \bibfield  {author} {\bibinfo {author} {\bibfnamefont {M.}~\bibnamefont
  {Weger}},\ }\bibfield  {title} {\bibinfo {title} {A model for the electronic
  band structure of {HMTSeF-TCNQ}},\ }\href@noop {} {\bibfield  {journal}
  {\bibinfo  {journal} {Solid State Commun.}\ }\textbf {\bibinfo {volume}
  {19}},\ \bibinfo {pages} {1149} (\bibinfo {year} {1976})}\BibitemShut
  {NoStop}%
\bibitem [{\citenamefont {Murata}\ \emph {et~al.}(2014)\citenamefont {Murata},
  \citenamefont {Fukumoto}, \citenamefont {Yokogawa}, \citenamefont {Takaoka},
  \citenamefont {Kang}, \citenamefont {Brooks}, \citenamefont {Graf},
  \citenamefont {Yoshino}, \citenamefont {Sasaki},\ and\ \citenamefont
  {Kato}}]{murata-lowtemp}%
  \BibitemOpen
  \bibfield  {author} {\bibinfo {author} {\bibfnamefont {K.}~\bibnamefont
  {Murata}}, \bibinfo {author} {\bibfnamefont {Y.}~\bibnamefont {Fukumoto}},
  \bibinfo {author} {\bibfnamefont {K.}~\bibnamefont {Yokogawa}}, \bibinfo
  {author} {\bibfnamefont {R.}~\bibnamefont {Takaoka}}, \bibinfo {author}
  {\bibfnamefont {W.}~\bibnamefont {Kang}}, \bibinfo {author} {\bibfnamefont
  {J.~S.}\ \bibnamefont {Brooks}}, \bibinfo {author} {\bibfnamefont
  {D.}~\bibnamefont {Graf}}, \bibinfo {author} {\bibfnamefont {H.}~\bibnamefont
  {Yoshino}}, \bibinfo {author} {\bibfnamefont {T.}~\bibnamefont {Sasaki}},\
  and\ \bibinfo {author} {\bibfnamefont {R.}~\bibnamefont {Kato}},\ }\bibfield
  {title} {\bibinfo {title} {Magnetic-field-induced phase transitions in the
  quasi-one-dimensional organic conductor {HMTSF-TCNQ}},\ }\href@noop {}
  {\bibfield  {journal} {\bibinfo  {journal} {Low Temp. Phys.}\ }\textbf
  {\bibinfo {volume} {40}},\ \bibinfo {pages} {371} (\bibinfo {year}
  {2014})}\BibitemShut {NoStop}%
\bibitem [{\citenamefont {Murata}\ \emph {et~al.}(2012)\citenamefont {Murata},
  \citenamefont {Kang}, \citenamefont {Masuda}, \citenamefont {Kuse},
  \citenamefont {Sasaki}, \citenamefont {Yokogawa}, \citenamefont {Yoshino},
  \citenamefont {Brooks}, \citenamefont {Choi}, \citenamefont {Kiswandhi},\
  and\ \citenamefont {Kato}}]{murata-physica}%
  \BibitemOpen
  \bibfield  {author} {\bibinfo {author} {\bibfnamefont {K.}~\bibnamefont
  {Murata}}, \bibinfo {author} {\bibfnamefont {W.}~\bibnamefont {Kang}},
  \bibinfo {author} {\bibfnamefont {K.}~\bibnamefont {Masuda}}, \bibinfo
  {author} {\bibfnamefont {T.}~\bibnamefont {Kuse}}, \bibinfo {author}
  {\bibfnamefont {T.}~\bibnamefont {Sasaki}}, \bibinfo {author} {\bibfnamefont
  {K.}~\bibnamefont {Yokogawa}}, \bibinfo {author} {\bibfnamefont
  {H.}~\bibnamefont {Yoshino}}, \bibinfo {author} {\bibfnamefont {J.~S.}\
  \bibnamefont {Brooks}}, \bibinfo {author} {\bibfnamefont {E.~S.}\
  \bibnamefont {Choi}}, \bibinfo {author} {\bibfnamefont {A.}~\bibnamefont
  {Kiswandhi}},\ and\ \bibinfo {author} {\bibfnamefont {R.}~\bibnamefont
  {Kato}},\ }\bibfield  {title} {\bibinfo {title} {Field-induced {CDW} in
  {HMTSF-TCNQ}},\ }\href@noop {} {\bibfield  {journal} {\bibinfo  {journal}
  {Physica B}\ }\textbf {\bibinfo {volume} {407}},\ \bibinfo {pages} {1927}
  (\bibinfo {year} {2012})}\BibitemShut {NoStop}%
\bibitem [{\citenamefont {Murata}\ \emph {et~al.}(2010)\citenamefont {Murata},
  \citenamefont {Yokogawa}, \citenamefont {Kobayashi}, \citenamefont {Masuda},
  \citenamefont {Sasaki}, \citenamefont {Seno}, \citenamefont
  {Rani~Tamilselvan}, \citenamefont {Yoshino}, \citenamefont {S.~Brooks},
  \citenamefont {J辿rome}, \citenamefont {Bechgaard}, \citenamefont {Uruichi},
  \citenamefont {Yakushi}, \citenamefont {Nogami},\ and\ \citenamefont
  {Kato}}]{murata-jpsj}%
  \BibitemOpen
  \bibfield  {author} {\bibinfo {author} {\bibfnamefont {K.}~\bibnamefont
  {Murata}}, \bibinfo {author} {\bibfnamefont {K.}~\bibnamefont {Yokogawa}},
  \bibinfo {author} {\bibfnamefont {K.}~\bibnamefont {Kobayashi}}, \bibinfo
  {author} {\bibfnamefont {K.}~\bibnamefont {Masuda}}, \bibinfo {author}
  {\bibfnamefont {T.}~\bibnamefont {Sasaki}}, \bibinfo {author} {\bibfnamefont
  {Y.}~\bibnamefont {Seno}}, \bibinfo {author} {\bibfnamefont {N.}~\bibnamefont
  {Rani~Tamilselvan}}, \bibinfo {author} {\bibfnamefont {H.}~\bibnamefont
  {Yoshino}}, \bibinfo {author} {\bibfnamefont {J.}~\bibnamefont {S.~Brooks}},
  \bibinfo {author} {\bibfnamefont {D.}~\bibnamefont {J辿rome}}, \bibinfo
  {author} {\bibfnamefont {K.}~\bibnamefont {Bechgaard}}, \bibinfo {author}
  {\bibfnamefont {M.}~\bibnamefont {Uruichi}}, \bibinfo {author} {\bibfnamefont
  {K.}~\bibnamefont {Yakushi}}, \bibinfo {author} {\bibfnamefont
  {Y.}~\bibnamefont {Nogami}},\ and\ \bibinfo {author} {\bibfnamefont
  {R.}~\bibnamefont {Kato}},\ }\bibfield  {title} {\bibinfo {title}
  {Field-induced successive phase transitions in the charge density wave
  organic conductor {HMTSF-TCNQ}},\ }\href@noop {} {\bibfield  {journal}
  {\bibinfo  {journal} {J. Phys. Soc. Jpn.}\ }\textbf {\bibinfo {volume}
  {79}},\ \bibinfo {pages} {103702} (\bibinfo {year} {2010})}\BibitemShut
  {NoStop}%
\bibitem [{\citenamefont {J\'{e}rome}\ and\ \citenamefont
  {Schulz}(1982)}]{jerome-review}%
  \BibitemOpen
  \bibfield  {author} {\bibinfo {author} {\bibfnamefont {D.}~\bibnamefont
  {J\'{e}rome}}\ and\ \bibinfo {author} {\bibfnamefont {H.~J.}\ \bibnamefont
  {Schulz}},\ }\bibfield  {title} {\bibinfo {title} {Organic conductors and
  superconductors},\ }\href@noop {} {\bibfield  {journal} {\bibinfo  {journal}
  {Advances in Physics}\ }\textbf {\bibinfo {volume} {31:4}},\ \bibinfo {pages}
  {299} (\bibinfo {year} {1982})}\BibitemShut {NoStop}%
\bibitem [{\citenamefont {J\'{e}rome}(2004)}]{jerome04}%
  \BibitemOpen
  \bibfield  {author} {\bibinfo {author} {\bibfnamefont {D.}~\bibnamefont
  {J\'{e}rome}},\ }\bibfield  {title} {\bibinfo {title} {Organic conductors: 
  from charge density wave {TTF−TCNQ} to superconducting
  ({TMTSF})$_2${PF}${}_6$},\ }\href@noop {} {\bibfield  {journal} {\bibinfo
  {journal} {Chemical Reviews}\ }\textbf {\bibinfo {volume} {104}},\ \bibinfo
  {pages} {5565} (\bibinfo {year} {2004})}\BibitemShut {NoStop}%
\bibitem [{\citenamefont {Phillips}\ \emph {et~al.}(1976)\citenamefont
  {Phillips}, \citenamefont {Kistenmacher}, \citenamefont {Bloch},\ and\
  \citenamefont {Cowan}}]{phillips}%
  \BibitemOpen
  \bibfield  {author} {\bibinfo {author} {\bibfnamefont {T.~E.}\ \bibnamefont
  {Phillips}}, \bibinfo {author} {\bibfnamefont {T.~J.}\ \bibnamefont
  {Kistenmacher}}, \bibinfo {author} {\bibfnamefont {A.~N.}\ \bibnamefont
  {Bloch}},\ and\ \bibinfo {author} {\bibfnamefont {D.~O.}\ \bibnamefont
  {Cowan}},\ }\href@noop {} {\bibfield  {journal} {\bibinfo  {journal} {J.
  Chem. Soc., Chem. Commun.}\ }\textbf {\bibinfo {volume} {66}},\ \bibinfo
  {pages} {334} (\bibinfo {year} {1976})}\BibitemShut {NoStop}%
\bibitem [{\citenamefont {Fukuyama}\ and\ \citenamefont
  {Kubo}(1970)}]{fukuyama-kubo}%
  \BibitemOpen
  \bibfield  {author} {\bibinfo {author} {\bibfnamefont {H.}~\bibnamefont
  {Fukuyama}}\ and\ \bibinfo {author} {\bibfnamefont {R.}~\bibnamefont
  {Kubo}},\ }\bibfield  {title} {\bibinfo {title} {Interband effects on
  magnetic susceptibility. {II}. diamagnetism of bismuth},\ }\href@noop {}
  {\bibfield  {journal} {\bibinfo  {journal} {J. Phys. Soc. Jpn.}\ }\textbf
  {\bibinfo {volume} {28}},\ \bibinfo {pages} {570} (\bibinfo {year}
  {1970})}\BibitemShut {NoStop}%
\bibitem [{\citenamefont {Fukuyama}(2007)}]{fukuyama07}%
  \BibitemOpen
  \bibfield  {author} {\bibinfo {author} {\bibfnamefont {H.}~\bibnamefont
  {Fukuyama}},\ }\bibfield  {title} {\bibinfo {title} {Anomalous orbital
  magnetism and {Hall} effect of massless fermions in two dimension},\
  }\href@noop {} {\bibfield  {journal} {\bibinfo  {journal} {J. Phys. Soc.
  Jpn.}\ }\textbf {\bibinfo {volume} {76}},\ \bibinfo {pages} {043711}
  (\bibinfo {year} {2007})}\BibitemShut {NoStop}%
\bibitem [{\citenamefont {Nakamura}(2007)}]{nakamura07}%
  \BibitemOpen
  \bibfield  {author} {\bibinfo {author} {\bibfnamefont {M.}~\bibnamefont
  {Nakamura}},\ }\bibfield  {title} {\bibinfo {title} {Orbital magnetism and
  transport phenomena in two-dimensional {Dirac} fermions in a weak magnetic
  field},\ }\href {https://doi.org/10.1103/PhysRevB.76.113301} {\bibfield
  {journal} {\bibinfo  {journal} {Phys. Rev. B}\ }\textbf {\bibinfo {volume}
  {76}},\ \bibinfo {pages} {113301} (\bibinfo {year} {2007})}\BibitemShut
  {NoStop}%
\bibitem [{\citenamefont {Koshino}\ and\ \citenamefont
  {Ando}(2010)}]{koshino-ando}%
  \BibitemOpen
  \bibfield  {author} {\bibinfo {author} {\bibfnamefont {M.}~\bibnamefont
  {Koshino}}\ and\ \bibinfo {author} {\bibfnamefont {T.}~\bibnamefont {Ando}},\
  }\bibfield  {title} {\bibinfo {title} {Anomalous orbital magnetism in
  {Dirac}-electron systems: Role of pseudospin paramagnetism},\ }\href@noop {}
  {\bibfield  {journal} {\bibinfo  {journal} {Phys. Rev. B}\ }\textbf {\bibinfo
  {volume} {81}},\ \bibinfo {pages} {195431} (\bibinfo {year}
  {2010})}\BibitemShut {NoStop}%
\bibitem [{\citenamefont {Raoux}\ \emph {et~al.}(2015)\citenamefont {Raoux},
  \citenamefont {Pi\'{e}chon}, \citenamefont {Fuchs},\ and\ \citenamefont
  {Montambaux}}]{raoux-piechon}%
  \BibitemOpen
  \bibfield  {author} {\bibinfo {author} {\bibfnamefont {A.}~\bibnamefont
  {Raoux}}, \bibinfo {author} {\bibfnamefont {F.}~\bibnamefont {Pi\'{e}chon}},
  \bibinfo {author} {\bibfnamefont {J.~N.}\ \bibnamefont {Fuchs}},\ and\
  \bibinfo {author} {\bibfnamefont {G.}~\bibnamefont {Montambaux}},\ }\bibfield
   {title} {\bibinfo {title} {Orbital magnetism in coupled-bands models},\
  }\href@noop {} {\bibfield  {journal} {\bibinfo  {journal} {Phys. Rev. B}\
  }\textbf {\bibinfo {volume} {91}},\ \bibinfo {pages} {085120} (\bibinfo
  {year} {2015})}\BibitemShut {NoStop}%
\bibitem [{\citenamefont {Fuseya}\ \emph {et~al.}(2015)\citenamefont {Fuseya},
  \citenamefont {Ogata},\ and\ \citenamefont {Fukuyama}}]{fuseya}%
  \BibitemOpen
  \bibfield  {author} {\bibinfo {author} {\bibfnamefont {Y.}~\bibnamefont
  {Fuseya}}, \bibinfo {author} {\bibfnamefont {M.}~\bibnamefont {Ogata}},\ and\
  \bibinfo {author} {\bibfnamefont {H.}~\bibnamefont {Fukuyama}},\ }\bibfield
  {title} {\bibinfo {title} {Transport properties and diamagnetism of {Dirac}
  electrons in bismuth},\ }\href@noop {} {\bibfield  {journal} {\bibinfo
  {journal} {J. Phys. Soc. Jpn.}\ }\textbf {\bibinfo {volume} {84}},\ \bibinfo
  {pages} {012001} (\bibinfo {year} {2015})}\BibitemShut {NoStop}%
\bibitem [{\citenamefont {Ozaki}\ and\ \citenamefont
  {Ogata}(2021)}]{ozaki-ogata}%
  \BibitemOpen
  \bibfield  {author} {\bibinfo {author} {\bibfnamefont {S.}~\bibnamefont
  {Ozaki}}\ and\ \bibinfo {author} {\bibfnamefont {M.}~\bibnamefont {Ogata}},\
  }\bibfield  {title} {\bibinfo {title} {Universal quantization of the magnetic
  susceptibility jump at a topological phase transition},\ }\href
  {https://doi.org/10.1103/PhysRevResearch.3.013058} {\bibfield  {journal}
  {\bibinfo  {journal} {Phys. Rev. Research}\ }\textbf {\bibinfo {volume}
  {3}},\ \bibinfo {pages} {013058} (\bibinfo {year} {2021})}\BibitemShut
  {NoStop}%
\bibitem [{\citenamefont {Fukuyama}(1971)}]{fukuyama71}%
  \BibitemOpen
  \bibfield  {author} {\bibinfo {author} {\bibfnamefont {H.}~\bibnamefont
  {Fukuyama}},\ }\bibfield  {title} {\bibinfo {title} {Theory of orbital
  magnetism of {Bloch} electrons: {Coulomb} interactions},\ }\href@noop {}
  {\bibfield  {journal} {\bibinfo  {journal} {Prog. Theor. Phys.}\ }\textbf
  {\bibinfo {volume} {45}},\ \bibinfo {pages} {704} (\bibinfo {year}
  {1971})}\BibitemShut {NoStop}%
\bibitem [{\citenamefont {Momma}\ and\ \citenamefont {Izumi}(2011)}]{vesta}%
  \BibitemOpen
  \bibfield  {author} {\bibinfo {author} {\bibfnamefont {K.}~\bibnamefont
  {Momma}}\ and\ \bibinfo {author} {\bibfnamefont {F.}~\bibnamefont {Izumi}},\
  }\bibfield  {title} {\bibinfo {title} {Vesta 3 for three-dimensional
  visualization of crystal, volumetric and morphology data},\ }\href@noop {}
  {\bibfield  {journal} {\bibinfo  {journal} {J. Appl. Crystallogr.}\ }\textbf
  {\bibinfo {volume} {44}},\ \bibinfo {pages} {1272} (\bibinfo {year}
  {2011})}\BibitemShut {NoStop}%
\bibitem [{\citenamefont {Hahn}\ \emph {et~al.}(1983)\citenamefont {Hahn},
  \citenamefont {Shmueli},\ and\ \citenamefont {Arthur}}]{hahn}%
  \BibitemOpen
  \bibfield  {author} {\bibinfo {author} {\bibfnamefont {T.}~\bibnamefont
  {Hahn}}, \bibinfo {author} {\bibfnamefont {U.}~\bibnamefont {Shmueli}},\ and\
  \bibinfo {author} {\bibfnamefont {J.~W.}\ \bibnamefont {Arthur}},\
  }\href@noop {} {\emph {\bibinfo {title} {International Tables for
  Crystallography}}},\ Vol.~\bibinfo {volume} {1}\ (\bibinfo  {publisher}
  {Reidel, Dordrecht},\ \bibinfo {year} {1983})\BibitemShut {NoStop}%
\bibitem [{\citenamefont {Cui}\ and\ \citenamefont {Kato}()}]{kato}%
  \BibitemOpen
  \bibfield  {author} {\bibinfo {author} {\bibfnamefont {H.}~\bibnamefont
  {Cui}}\ and\ \bibinfo {author} {\bibfnamefont {R.}~\bibnamefont {Kato}},\
  }\bibinfo {note} {private communication}\BibitemShut {NoStop}%
\bibitem [{\citenamefont {Giannozzi}\ \emph {et~al.}(2009)\citenamefont
  {Giannozzi}, \citenamefont {Baroni}, \citenamefont {Bonini}, \citenamefont
  {Calandra}, \citenamefont {Car}, \citenamefont {Cavazzoni}, \citenamefont
  {Ceresoli}, \citenamefont {Chiarotti}, \citenamefont {Cococcioni},\ and\
  \citenamefont {et~al.}}]{qe}%
  \BibitemOpen
  \bibfield  {author} {\bibinfo {author} {\bibfnamefont {P.}~\bibnamefont
  {Giannozzi}}, \bibinfo {author} {\bibfnamefont {S.}~\bibnamefont {Baroni}},
  \bibinfo {author} {\bibfnamefont {N.}~\bibnamefont {Bonini}}, \bibinfo
  {author} {\bibfnamefont {M.}~\bibnamefont {Calandra}}, \bibinfo {author}
  {\bibfnamefont {R.}~\bibnamefont {Car}}, \bibinfo {author} {\bibfnamefont
  {C.}~\bibnamefont {Cavazzoni}}, \bibinfo {author} {\bibfnamefont
  {D.}~\bibnamefont {Ceresoli}}, \bibinfo {author} {\bibfnamefont {G.~L.}\
  \bibnamefont {Chiarotti}}, \bibinfo {author} {\bibfnamefont {M.}~\bibnamefont
  {Cococcioni}},\ and\ \bibinfo {author} {\bibfnamefont {I.~D.}\ \bibnamefont
  {et~al.}},\ }\href@noop {} {\bibfield  {journal} {\bibinfo  {journal} {J.
  Phys. : Condens. Matter}\ }\textbf {\bibinfo {volume} {21}},\ \bibinfo
  {pages} {395502} (\bibinfo {year} {2009})}\BibitemShut {NoStop}%
\bibitem [{\citenamefont {Hohenberg}\ and\ \citenamefont
  {Kohn}(1964)}]{hohenberg-kohn}%
  \BibitemOpen
  \bibfield  {author} {\bibinfo {author} {\bibfnamefont {P.}~\bibnamefont
  {Hohenberg}}\ and\ \bibinfo {author} {\bibfnamefont {W.}~\bibnamefont
  {Kohn}},\ }\bibfield  {title} {\bibinfo {title} {Inhomogeneous electron
  gas},\ }\href@noop {} {\bibfield  {journal} {\bibinfo  {journal} {Phys.
  Rev.}\ }\textbf {\bibinfo {volume} {136}},\ \bibinfo {pages} {B864} (\bibinfo
  {year} {1964})}\BibitemShut {NoStop}%
\bibitem [{\citenamefont {Kohn}\ and\ \citenamefont {Sham}(1965)}]{kohn-sham}%
  \BibitemOpen
  \bibfield  {author} {\bibinfo {author} {\bibfnamefont {W.}~\bibnamefont
  {Kohn}}\ and\ \bibinfo {author} {\bibfnamefont {L.~J.}\ \bibnamefont
  {Sham}},\ }\bibfield  {title} {\bibinfo {title} {Self-consistent equations
  including exchange and correlation effects},\ }\href@noop {} {\bibfield
  {journal} {\bibinfo  {journal} {Phys. Rev.}\ }\textbf {\bibinfo {volume}
  {140}},\ \bibinfo {pages} {A1133} (\bibinfo {year} {1965})}\BibitemShut
  {NoStop}%
\bibitem [{\citenamefont {Perdew}\ \emph {et~al.}(1996)\citenamefont {Perdew},
  \citenamefont {Burke},\ and\ \citenamefont {Ernzerhof}}]{pbe96}%
  \BibitemOpen
  \bibfield  {author} {\bibinfo {author} {\bibfnamefont {J.~P.}\ \bibnamefont
  {Perdew}}, \bibinfo {author} {\bibfnamefont {K.}~\bibnamefont {Burke}},\ and\
  \bibinfo {author} {\bibfnamefont {M.}~\bibnamefont {Ernzerhof}},\ }\bibfield
  {title} {\bibinfo {title} {Generalized gradient approximation made simple},\
  }\href@noop {} {\bibfield  {journal} {\bibinfo  {journal} {Phys. Rev. Lett}\
  }\textbf {\bibinfo {volume} {77}},\ \bibinfo {pages} {3865} (\bibinfo {year}
  {1996})}\BibitemShut {NoStop}%
\bibitem [{\citenamefont {Fu}\ and\ \citenamefont {Kane}(2007)}]{fu-kane}%
  \BibitemOpen
  \bibfield  {author} {\bibinfo {author} {\bibfnamefont {L.}~\bibnamefont
  {Fu}}\ and\ \bibinfo {author} {\bibfnamefont {C.~L.}\ \bibnamefont {Kane}},\
  }\bibfield  {title} {\bibinfo {title} {Topological insulators with inversion
  symmetry},\ }\href {https://doi.org/10.1103/PhysRevB.76.045302} {\bibfield
  {journal} {\bibinfo  {journal} {Phys. Rev. B}\ }\textbf {\bibinfo {volume}
  {76}},\ \bibinfo {pages} {045302} (\bibinfo {year} {2007})}\BibitemShut
  {NoStop}%
\bibitem [{\citenamefont {Po}\ \emph {et~al.}(2017)\citenamefont {Po},
  \citenamefont {Vishwanath},\ and\ \citenamefont {Watanabe}}]{pvw}%
  \BibitemOpen
  \bibfield  {author} {\bibinfo {author} {\bibfnamefont {H.}~\bibnamefont
  {Po}}, \bibinfo {author} {\bibfnamefont {A.}~\bibnamefont {Vishwanath}},\
  and\ \bibinfo {author} {\bibfnamefont {H.}~\bibnamefont {Watanabe}},\
  }\bibfield  {title} {\bibinfo {title} {Symmetry-based indicators of band
  topology in the 230 space groups},\ }\href@noop {} {\bibfield  {journal}
  {\bibinfo  {journal} {Nat. Commun.}\ }\textbf {\bibinfo {volume} {8}},\
  \bibinfo {pages} {50} (\bibinfo {year} {2017})}\BibitemShut {NoStop}%
\bibitem [{\citenamefont {Song}\ \emph
  {et~al.}(2018{\natexlab{a}})\citenamefont {Song}, \citenamefont {Zhang},
  \citenamefont {Fang},\ and\ \citenamefont {Fang}}]{song-natcom}%
  \BibitemOpen
  \bibfield  {author} {\bibinfo {author} {\bibfnamefont {Z.}~\bibnamefont
  {Song}}, \bibinfo {author} {\bibfnamefont {T.}~\bibnamefont {Zhang}},
  \bibinfo {author} {\bibfnamefont {Z.}~\bibnamefont {Fang}},\ and\ \bibinfo
  {author} {\bibfnamefont {C.}~\bibnamefont {Fang}},\ }\bibfield  {title}
  {\bibinfo {title} {Quantative mappings between symmetry and topology in
  solids},\ }\href@noop {} {\bibfield  {journal} {\bibinfo  {journal} {Nat.
  Commun.}\ }\textbf {\bibinfo {volume} {9}},\ \bibinfo {pages} {3530}
  (\bibinfo {year} {2018}{\natexlab{a}})}\BibitemShut {NoStop}%
\bibitem [{\citenamefont {Kim}\ \emph {et~al.}(2015)\citenamefont {Kim},
  \citenamefont {Wieder}, \citenamefont {Kane},\ and\ \citenamefont
  {Rappe}}]{youngkuk}%
  \BibitemOpen
  \bibfield  {author} {\bibinfo {author} {\bibfnamefont {Y.}~\bibnamefont
  {Kim}}, \bibinfo {author} {\bibfnamefont {B.~J.}\ \bibnamefont {Wieder}},
  \bibinfo {author} {\bibfnamefont {C.~L.}\ \bibnamefont {Kane}},\ and\
  \bibinfo {author} {\bibfnamefont {A.~M.}\ \bibnamefont {Rappe}},\ }\bibfield
  {title} {\bibinfo {title} {Dirac line nodes in inversion-symmetric
  crystals},\ }\href {https://doi.org/10.1103/PhysRevLett.115.036806}
  {\bibfield  {journal} {\bibinfo  {journal} {Phys. Rev. Lett.}\ }\textbf
  {\bibinfo {volume} {115}},\ \bibinfo {pages} {036806} (\bibinfo {year}
  {2015})}\BibitemShut {NoStop}%
\bibitem [{\citenamefont {Song}\ \emph
  {et~al.}(2018{\natexlab{b}})\citenamefont {Song}, \citenamefont {Zhang},\
  and\ \citenamefont {Fang}}]{song-prx}%
  \BibitemOpen
  \bibfield  {author} {\bibinfo {author} {\bibfnamefont {Z.}~\bibnamefont
  {Song}}, \bibinfo {author} {\bibfnamefont {T.}~\bibnamefont {Zhang}},\ and\
  \bibinfo {author} {\bibfnamefont {C.}~\bibnamefont {Fang}},\ }\bibfield
  {title} {\bibinfo {title} {Diagnosis for nonmagnetic topological semimetals
  in the absence of spin-orbital coupling},\ }\href@noop {} {\bibfield
  {journal} {\bibinfo  {journal} {Phys. Rev. X}\ }\textbf {\bibinfo {volume}
  {8}},\ \bibinfo {pages} {031069} (\bibinfo {year}
  {2018}{\natexlab{b}})}\BibitemShut {NoStop}%
\bibitem [{\citenamefont {Slater}\ and\ \citenamefont {Koster}(1954)}]{sk}%
  \BibitemOpen
  \bibfield  {author} {\bibinfo {author} {\bibfnamefont {J.~C.}\ \bibnamefont
  {Slater}}\ and\ \bibinfo {author} {\bibfnamefont {G.~F.}\ \bibnamefont
  {Koster}},\ }\bibfield  {title} {\bibinfo {title} {Simplified {LCAO} method
  for the periodic potential problem},\ }\href@noop {} {\bibfield  {journal}
  {\bibinfo  {journal} {Phys. Rev.}\ }\textbf {\bibinfo {volume} {94}},\
  \bibinfo {pages} {1498} (\bibinfo {year} {1954})}\BibitemShut {NoStop}%
\bibitem [{\citenamefont {Mostofi}\ \emph {et~al.}(2008)\citenamefont
  {Mostofi}, \citenamefont {Yates}, \citenamefont {Lee}, \citenamefont {Souza},
  \citenamefont {Vanderbilt},\ and\ \citenamefont {Marzari}}]{wannier90}%
  \BibitemOpen
  \bibfield  {author} {\bibinfo {author} {\bibfnamefont {A.~A.}\ \bibnamefont
  {Mostofi}}, \bibinfo {author} {\bibfnamefont {J.~R.}\ \bibnamefont {Yates}},
  \bibinfo {author} {\bibfnamefont {Y.-S.}\ \bibnamefont {Lee}}, \bibinfo
  {author} {\bibfnamefont {I.}~\bibnamefont {Souza}}, \bibinfo {author}
  {\bibfnamefont {D.}~\bibnamefont {Vanderbilt}},\ and\ \bibinfo {author}
  {\bibfnamefont {N.}~\bibnamefont {Marzari}},\ }\bibfield  {title} {\bibinfo
  {title} {An updated version of wannier90: A tool for obtaining
  maximally-localised {Wannier} functions},\ }\href@noop {} {\bibfield
  {journal} {\bibinfo  {journal} {Comput. Phys. Commun.}\ }\textbf {\bibinfo
  {volume} {77}},\ \bibinfo {pages} {3865} (\bibinfo {year}
  {2008})}\BibitemShut {NoStop}%
\bibitem [{\citenamefont {Ravy}\ \emph {et~al.}(2005)\citenamefont {Ravy},
  \citenamefont {Launois}, \citenamefont {Moret},\ and\ \citenamefont
  {Pouget}}]{ravy}%
  \BibitemOpen
  \bibfield  {author} {\bibinfo {author} {\bibfnamefont {S.}~\bibnamefont
  {Ravy}}, \bibinfo {author} {\bibfnamefont {P.}~\bibnamefont {Launois}},
  \bibinfo {author} {\bibfnamefont {R.}~\bibnamefont {Moret}},\ and\ \bibinfo
  {author} {\bibfnamefont {J.-P.}\ \bibnamefont {Pouget}},\ }\bibfield  {title}
  {\bibinfo {title} {Case studies of molecular disorder},\ }\href@noop {}
  {\bibfield  {journal} {\bibinfo  {journal} {Z. Kristallogr.}\ }\textbf
  {\bibinfo {volume} {220}},\ \bibinfo {pages} {1059} (\bibinfo {year}
  {2005})}\BibitemShut {NoStop}%
\bibitem [{\citenamefont {Fro\"{o}hlich}(1954)}]{frorich}%
  \BibitemOpen
  \bibfield  {author} {\bibinfo {author} {\bibfnamefont {H.}~\bibnamefont
  {Fro\"{o}hlich}},\ }\bibfield  {title} {\bibinfo {title} {On the theory of
  superconductivity: the one-dimensional case},\ }\href@noop {} {\bibfield
  {journal} {\bibinfo  {journal} {Proc. R. Soc. Lond. A}\ }\textbf {\bibinfo
  {volume} {223}},\ \bibinfo {pages} {296} (\bibinfo {year}
  {1954})}\BibitemShut {NoStop}%
\bibitem [{\citenamefont {Kuper}(1955)}]{kuper}%
  \BibitemOpen
  \bibfield  {author} {\bibinfo {author} {\bibfnamefont {C.~G.}\ \bibnamefont
  {Kuper}},\ }\bibfield  {title} {\bibinfo {title} {The thermal decomposition
  of ammonium perchlorate {II}. the kinetics of the decomposition, the effect
  of particle size, and discussion of results},\ }\href@noop {} {\bibfield
  {journal} {\bibinfo  {journal} {Proc. R. Soc. Lond. A}\ }\textbf {\bibinfo
  {volume} {227}},\ \bibinfo {pages} {214} (\bibinfo {year}
  {1955})}\BibitemShut {NoStop}%
\bibitem [{\citenamefont {Rice}\ and\ \citenamefont
  {Str\"{a}ssler}(1973)}]{rice-strassler}%
  \BibitemOpen
  \bibfield  {author} {\bibinfo {author} {\bibfnamefont {M.~J.}\ \bibnamefont
  {Rice}}\ and\ \bibinfo {author} {\bibfnamefont {S.}~\bibnamefont
  {Str\"{a}ssler}},\ }\bibfield  {title} {\bibinfo {title} {Theory of a
  quasi-one-dimensional band-conductor},\ }\href@noop {} {\bibfield  {journal}
  {\bibinfo  {journal} {Solid State Commun.}\ }\textbf {\bibinfo {volume}
  {13}},\ \bibinfo {pages} {125} (\bibinfo {year} {1973})}\BibitemShut
  {NoStop}%
\bibitem [{\citenamefont {Moriya}(1963)}]{moriya}%
  \BibitemOpen
  \bibfield  {author} {\bibinfo {author} {\bibfnamefont {T.}~\bibnamefont
  {Moriya}},\ }\bibfield  {title} {\bibinfo {title} {The effect of
  electron-electron interaction on the nuclear spin relaxation in metals},\
  }\href@noop {} {\bibfield  {journal} {\bibinfo  {journal} {J. Phys. Soc.
  Jpn.}\ }\textbf {\bibinfo {volume} {18}},\ \bibinfo {pages} {516} (\bibinfo
  {year} {1963})}\BibitemShut {NoStop}%
\bibitem [{\citenamefont {Suzumura}(1989)}]{suzumura89}%
  \BibitemOpen
  \bibfield  {author} {\bibinfo {author} {\bibfnamefont {Y.}~\bibnamefont
  {Suzumura}},\ }\bibfield  {title} {\bibinfo {title} {{NMR} relaxation rate of
  impure anisotropic quasi-one-dimensional superconductors},\ }\href@noop {}
  {\bibfield  {journal} {\bibinfo  {journal} {J. Phys. Soc. Jpn.}\ }\textbf
  {\bibinfo {volume} {58}},\ \bibinfo {pages} {2642} (\bibinfo {year}
  {1989})}\BibitemShut {NoStop}%
\bibitem [{\citenamefont {Katayama}\ \emph {et~al.}(2009)\citenamefont
  {Katayama}, \citenamefont {Kobayashi},\ and\ \citenamefont
  {Suzumura}}]{katayama}%
  \BibitemOpen
  \bibfield  {author} {\bibinfo {author} {\bibfnamefont {S.}~\bibnamefont
  {Katayama}}, \bibinfo {author} {\bibfnamefont {A.}~\bibnamefont
  {Kobayashi}},\ and\ \bibinfo {author} {\bibfnamefont {Y.}~\bibnamefont
  {Suzumura}},\ }\bibfield  {title} {\bibinfo {title} {Electronic properties
  close to {Dirac} cone in two-dimensional organic conductor
  $\alpha$-({BEDT-TTF})$_2${I}$_3$},\ }\href@noop {} {\bibfield  {journal}
  {\bibinfo  {journal} {Eur. Phys. J. B}\ }\textbf {\bibinfo {volume} {67}},\
  \bibinfo {pages} {139} (\bibinfo {year} {2009})}\BibitemShut {NoStop}%
\bibitem [{\citenamefont {Fetter}\ and\ \citenamefont
  {Walecka}(2003)}]{fetterwalecka}%
  \BibitemOpen
  \bibfield  {author} {\bibinfo {author} {\bibfnamefont {A.~L.}\ \bibnamefont
  {Fetter}}\ and\ \bibinfo {author} {\bibfnamefont {J.~D.}\ \bibnamefont
  {Walecka}},\ }\href@noop {} {\emph {\bibinfo {title} {Quantum theory of
  many-particle systems}}}\ (\bibinfo  {publisher} {Dover},\ \bibinfo {year}
  {2003})\BibitemShut {NoStop}%
\bibitem [{\citenamefont {Cooper}\ \emph {et~al.}(1976)\citenamefont {Cooper},
  \citenamefont {Weger}, \citenamefont {Jérome}, \citenamefont {Lefur},
  \citenamefont {Bechgaard}, \citenamefont {Bloch},\ and\ \citenamefont
  {Cowan}}]{cooper76}%
  \BibitemOpen
  \bibfield  {author} {\bibinfo {author} {\bibfnamefont {J.}~\bibnamefont
  {Cooper}}, \bibinfo {author} {\bibfnamefont {M.}~\bibnamefont {Weger}},
  \bibinfo {author} {\bibfnamefont {D.}~\bibnamefont {Jérome}}, \bibinfo
  {author} {\bibfnamefont {D.}~\bibnamefont {Lefur}}, \bibinfo {author}
  {\bibfnamefont {K.}~\bibnamefont {Bechgaard}}, \bibinfo {author}
  {\bibfnamefont {A.}~\bibnamefont {Bloch}},\ and\ \bibinfo {author}
  {\bibfnamefont {D.}~\bibnamefont {Cowan}},\ }\bibfield  {title} {\bibinfo
  {title} {Semi-metallic behaviour of {HMTSF-TCNQ} at low temperatures under
  pressure},\ }\href
  {https://doi.org/https://doi.org/10.1016/0038-1098(76)90911-X} {\bibfield
  {journal} {\bibinfo  {journal} {Solid State Communications}\ }\textbf
  {\bibinfo {volume} {19}},\ \bibinfo {pages} {749} (\bibinfo {year}
  {1976})}\BibitemShut {NoStop}%
\bibitem [{\citenamefont {Ogata}(2016)}]{ogata3}%
  \BibitemOpen
  \bibfield  {author} {\bibinfo {author} {\bibfnamefont {M.}~\bibnamefont
  {Ogata}},\ }\bibfield  {title} {\bibinfo {title} {Orbital magnetism of
  {Bloch} electrons: {III}. application to graphene},\ }\href@noop {}
  {\bibfield  {journal} {\bibinfo  {journal} {J. Phys. Soc. Jpn.}\ }\textbf
  {\bibinfo {volume} {85}},\ \bibinfo {pages} {104708} (\bibinfo {year}
  {2016})}\BibitemShut {NoStop}%
\bibitem [{\citenamefont {McClure}(1956)}]{mcclure}%
  \BibitemOpen
  \bibfield  {author} {\bibinfo {author} {\bibfnamefont {J.~W.}\ \bibnamefont
  {McClure}},\ }\bibfield  {title} {\bibinfo {title} {Diamagnetism of
  graphite},\ }\href@noop {} {\bibfield  {journal} {\bibinfo  {journal} {Phys.
  Rev.}\ }\textbf {\bibinfo {volume} {104}},\ \bibinfo {pages} {666} (\bibinfo
  {year} {1956})}\BibitemShut {NoStop}%
\bibitem [{\citenamefont {Tateishi}\ \emph {et~al.}()\citenamefont {Tateishi},
  \citenamefont {K\"{o}nye}, \citenamefont {Matsuura},\ and\ \citenamefont
  {Ogata}}]{tateishi}%
  \BibitemOpen
  \bibfield  {author} {\bibinfo {author} {\bibfnamefont {I.}~\bibnamefont
  {Tateishi}}, \bibinfo {author} {\bibfnamefont {V.}~\bibnamefont {K\"{o}nye}},
  \bibinfo {author} {\bibfnamefont {H.}~\bibnamefont {Matsuura}},\ and\
  \bibinfo {author} {\bibfnamefont {M.}~\bibnamefont {Ogata}},\ }\bibfield
  {title} {\bibinfo {title} {Characteristic singular behaviors of nodal line
  materials emerging in orbital magnetic susceptibility and {Hall}
  conductivity},\ }\Eprint {https://arxiv.org/abs/2103.05591}
  {arXiv:2103.05591} \BibitemShut {NoStop}%
\end{thebibliography}%

\end{document}